\documentclass[acmlarge]{acmart}

\usepackage{xspace}
\usepackage{subcaption}
\usepackage{graphicx} 
\usepackage{multirow}
\usepackage{makecell}
\usepackage[normalem]{ulem}
\usepackage{enumitem}
\usepackage{wrapfig}
\usepackage{rotating}
\usepackage[linesnumbered,ruled,vlined, noend]{algorithm2e}

\begin{document}

\title{GroupBeaMR: Analyzing Collaborative Group Behavior in Mixed Reality Through Passive Sensing and Sociometry}

\author{Diana Romero}
\authornote{Both authors contributed equally to this work.}
\email{dgromer1@uci.edu}
\affiliation{
  \institution{University of California, Irvine}
  \city{Irvine}
  \state{CA}
  \country{USA}
}

\author{Yasra Chandio}
\authornotemark[1]
\email{ychandio@umass.edu}
\affiliation{
  \institution{University of Massachusetts Amherst}
  \city{Amherst}
  \country{USA}}

\author{Fatima Anwar}
\email{fanwar@umass.edu}
\affiliation{
  \institution{University of Massachusetts Amherst}
  \city{Amherst}
  \country{USA}
}

\author{Salma Elmalaki}
\affiliation{
 \institution{University of California, Irvine}
 \city{Irvine}
 \state{CA}
 \country{USA}}
 \email{salma.elmalaki@uci.edu}

\renewcommand{\shortauthors}{Romero, Chandio, et al.}

\newcommand{\sysname}{\textbf{GroupBeaMR}\xspace}

\begin{abstract}
  Understanding group behavior is crucial for enhancing collaboration and productivity in mixed reality (MR). This paper introduces a framework for \uline{group} \uline{be}havior \uline{a}nalysis in \uline{MR}, or \textbf{\sysname} for short for analyzing group behavior in MR. \sysname leverages MR headsets' sensors to analyze group behavior through conversation, shared attention, and proximity, identifying \texttt{cohesive, fragmented, and competitive} interaction patterns. Using social network analysis, \sysname provides quantitative assessments of group dynamics, offering insights into collaboration structures. A user study with 48 participants across 12 groups validates the framework's ability to distinguish interaction patterns in MR environments. Our analyses show that group behavior is independent of task performance, emphasizing the significance of social interaction patterns.
  \sysname’s group type assignments indicate that sensor-based assessments in MR can provide meaningful insights into collaborative experiences, supporting the design of systems that adapt and optimize group behaviors.
\end{abstract}


\maketitle

\section{Introduction}\label{sec:intro}
Effective collaboration hinges on understanding group behavior, a complex interplay of communication patterns, decision-making, social norms, and power dynamics~\cite{leavitt1951some,kerr2004group,lapinski2005explication,genccer2019group}.  These dynamics are influenced by factors such as group size, composition, and task complexity~\cite{hackman1970effects,halfhill2005group,higgs2005influence}, ultimately impacting team productivity, creativity, and performance~\cite{stogdill1972group, paulus2000groups, podsakoff1997organizational}. While traditional research has explored group behavior in physical settings, the emergence of immersive technologies like Virtual Reality (VR), Augmented Reality (AR), and Mixed Reality (MR)\footnote{VR immerses users in a digital world, AR overlays digital information onto the real world, and MR blends the two, allowing digital and physical objects to interact in real-time.} offers new frontiers for collaborative work.  MR, in particular, presents a unique opportunity by seamlessly integrating digital objects into the physical environment, enabling interaction as if they were physically present.  This distinct blend of real and virtual distinguishes MR from VR (fully immersive) and AR (digital overlay), creating a unique collaborative landscape with significant implications for group dynamics.

However, understanding group behavior in MR presents significant challenges.  The complex interplay of real and virtual elements introduces novel variables like spatial awareness~\cite{luoWhereShouldWe2022,mcgeeComparisonSpatialVisualization2024}, presence~\cite{irlittiVolumetricMixedReality2023, demarbreInvestigatingPresenceRendering2024, tranSurveyMeasuringPresence2024}, and multimodal communication~\cite{szczurekMultimodalMultiUserMixed2023, mathisMRDrivenFutureRealities2024, guoEnhancingDigitalInheritance2024}, which can significantly influence how individuals interact and collaborate.  Moreover, the intricate nature of group interactions in MR, involving inter-, intra-, and multi-human variability~\cite{zhao2024fairo, elmalaki2021fair}, requires robust methods for capturing and analyzing these complex dynamics.  While research in VR and AR has provided valuable insights into group behavior in virtual and augmented environments~\cite{hanUnderstandingGroupBehavior2022,moustafaLongitudinalStudySmall2018,letarnecImprovingCollaborativeLearning2023,abdullahVideoconferenceEmbodiedVR2021,steedLeadershipCollaborationShared1999,hwang2013analysis, slater2000small}, and pervasive sensing technologies like Sociometric Badges~\cite{kimSociometricBadgesUsing2012} have advanced our understanding through data-driven insights~\cite{zhang_teamsense_2018, chaffinPromisePerilsWearable2017, olguinCapturingIndividualGroup2009}, translating these findings to the unique context of MR is not straightforward.  Existing tools and systems, while valuable, are insufficient to capture the nuanced interplay of real and virtual interactions in MR collaboration~\cite{ensRevisitingCollaborationMixed2019}. This gap is particularly critical given the rapid growth of the MR market, projected to double from \$1.4 billion in 2023 to 2030~\cite{llcMixedRealityMarket2023}, underscoring the increasing importance of effective MR collaboration. For example, as illustrated in ~\autoref{fig:showcase}, a collaborative MR setup showing participants engaging in an interactive image sorting game highlights how MR environments can foster both interactive and immersive team experiences. 

Therefore, this paper addresses the crucial need for a deeper understanding of group behavior in MR by answering the following key questions: (1) How can the sensory systems in MR headsets effectively capture group behavior during collaborative tasks? (2) What algorithms can process and interpret the data to infer group behavior? (3) Can we correlate group behavior with task-related performance metrics?  To address these questions, we introduce \sysname, a novel framework for \uline{g}roup \uline{b}ehavior \uline{a}nalysis in \uline{MR}.  Our contributions are as follows:

\begin{figure}[t]
\centering

\includegraphics[width=0.6\textwidth]{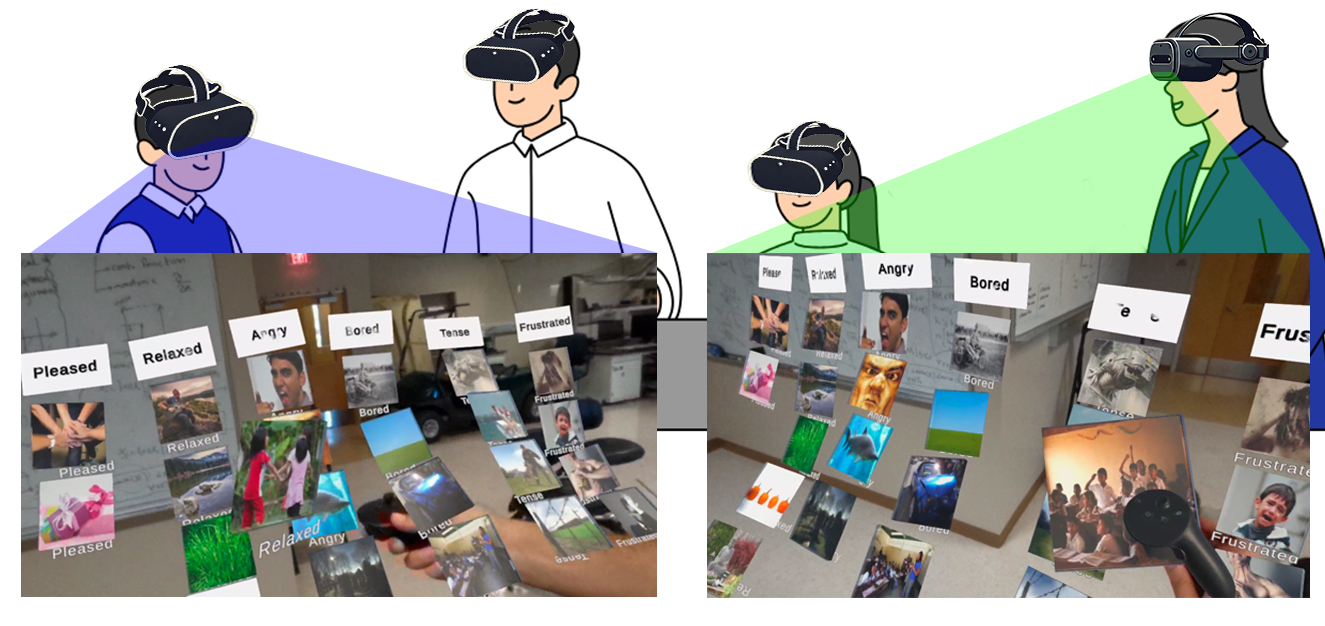}
~
\includegraphics[width=0.28\textwidth]{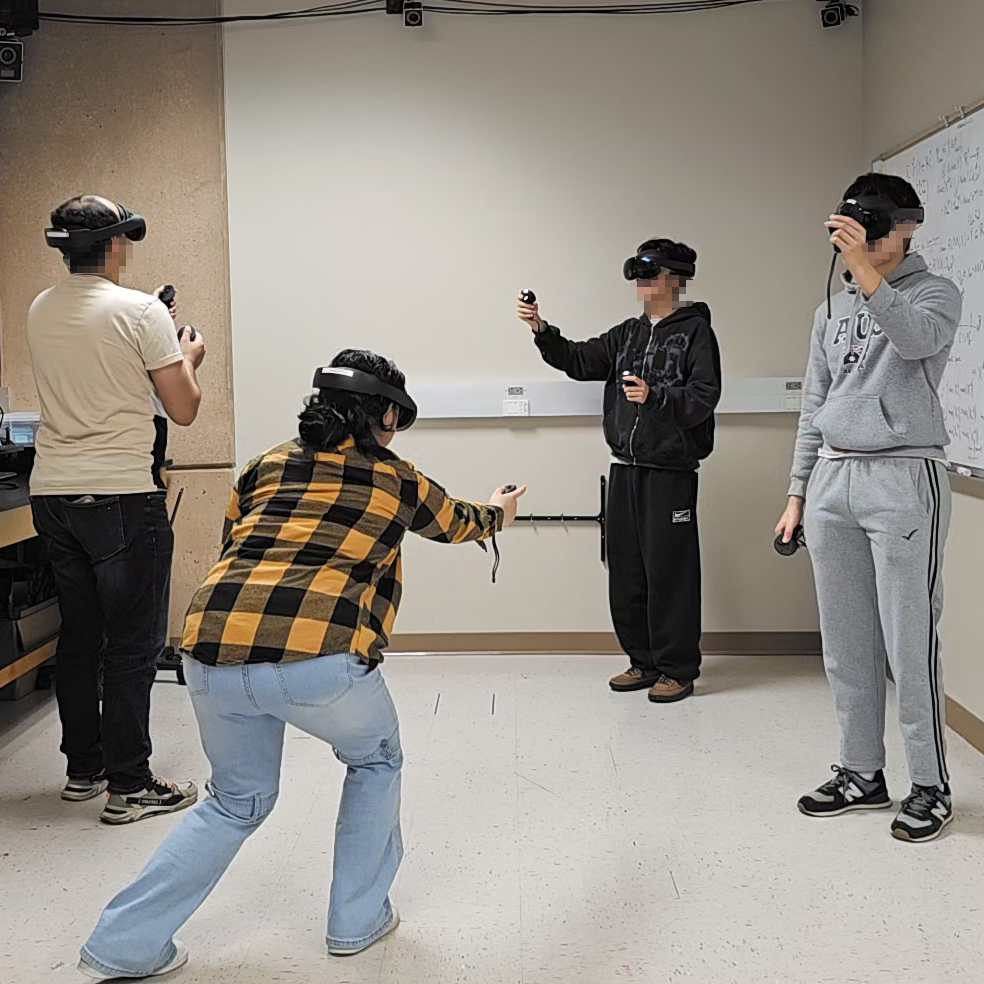}
\vspace{-0.3cm}
\caption{Illustration of the Collaborative MR Image Sorting Application. Left and Middle: First-person perspective of the application interface as experienced by two participants. Right: Third-person view showing participants interacting with the application in a shared MR environment.} \label{fig:showcase}
\end{figure}

\begin{enumerate}[noitemsep, topsep=0pt, leftmargin=*]
\item \textbf{Passive Sensing for Group Behavior Detection:}
\sysname is designed for unobtrusive data collection, preserving the natural flow of group interactions and the overall MR experience.  It captures a rich range of sensor modalities available in state-of-the-art MR headsets, providing a holistic understanding of group behavior.  Furthermore, \sysname operates ubiquitously, acquiring data efficiently without human intervention. This addresses the need for capturing the complex dynamics of MR collaboration without disrupting the experience.

\item \textbf{Interpretation of Group Behavior} 
We developed a systematic framework grounded in social network analysis to interpret and classify group behavior. This framework analyzes patterns of interaction and influence within the group, providing crucial insights by exploiting sociometric data and network analysis. This directly addresses the challenge of interpreting the rich data captured by MR sensors.

\item \textbf{Application of \sysname on Real-World User Study}: 
We validated \sysname through a comprehensive user study with 44 participants, examining its efficacy in capturing and interpreting group behavior within a real-world collaborative MR environment. This provides empirical evidence of the system's practical applicability.

\item \textbf{Correlation Analysis Between Task Performance and Group Behavior}: 
We analyzed quantitative and qualitative data from a collaborative MR task to assess the impact of group behavior on task outcomes. This analysis provides empirical insights into how specific behavioral and interaction patterns correlate with performance metrics, directly addressing the question of how group behavior affects collaborative success in MR.

\end{enumerate}

This paper is organized as follows: We cover background and related work in the literature in~\autoref{sec:related}. The proposed \sysname framework is described in~\autoref{sec:framework} and ~\autoref{sec:analysismodule}. Our user study and experimental setup are discussed in~\autoref{sec:user}. Our evaluation for \sysname is described in~\autoref{sec:results}. We provide further discussion implications of our work, limitations on our work, and future work in~\autoref{sec:discuss} and~\autoref{sec:limitations}. We finally conclude in~\autoref{sec:conclusion}.

\subsection{Observations and Impact}
Our results highlight that task performance does not directly shape group behavior. Instead, group dynamics emerge independently of measured performance metrics, suggesting that social interactions are influenced by cognitive, behavioral, and contextual factors beyond task outcomes. This finding underscores the complexity of collaborative MR environments, where engagement, communication, and coordination operate independently of traditional task efficiency measures. Our \sysname system is particularly valuable in this scenario, as it enables unobtrusive and data-driven analysis of group interactions in MR, capturing real-time social dynamics without interfering with user experience. Unlike traditional observation-based approaches, \emph{\sysname leverages sociometric sensing, network analysis, and interaction modeling to provide a structured framework for characterizing group behavior}. This capability is useful for identifying behavioral patterns, assessing engagement, and refining collaborative MR system designs to optimize teamwork beyond mere task performance metrics. By enabling data-driven behavioral profiling, \sysname supports adaptive MR system design, facilitating personalized interventions that enhance presence, engagement, and collaboration. The system provides a foundation for future studies exploring additional factors such as cognitive load, personality traits, and spatial adaptation, which may further explain how social interactions shape and evolve in MR.

\section{Background and Related Work}\label{sec:related}

\subsection{Group and Individual Behavior in Virtual and Mixed Reality Environments}

The study of group behavior in virtual environments has been extensive, demonstrating the impact of VR on social presence, interpersonal dynamics, and collaboration ~\cite{bailenson2008use, hauber2006spatiality, gatica2005detecting}. These studies highlight the interconnectedness of user actions and group dynamics in virtual settings. Theoretical frameworks such as the Social Identity Theory ~\cite{tajfel2004social} and the Proteus Effect ~\cite{yee2007proteus} provide insights into how user behavior in virtual environments reflects underlying group dynamics, influenced by group identity, social norms, and peer interaction.
MR introduces new variables to this field by blending physical and digital elements, necessitating a deeper exploration of group behavior in these hybrid environments. Research suggests that eye gaze behaviors differ between AR/MR head-mounted devices (HMD) conditions and traditional face-to-face interactions~\cite{prytzImportanceEyecontactCollaboration2010}. However, most research on human behavior in head-mounted MR displays has primarily focused on individual interactions with the technology, exploring how to design and display virtual elements effectively for everyday activities such as walking~\cite{changExperienceDesigningAugmented2024, lagesWalkingAdaptiveAugmented2019,leeExploringEffectsAugmented2023}, biking~\cite{koschNotiBikeAssessingTarget2022, chatterjeeSmartHelmMultimodalDetection2021, matviienkoBikeARUnderstandingCyclists2022}, and driving~\cite{stefanidiAugmentedRealityMove2024, ruschDirectingDriverAttention2013, medenicaAugmentedRealityVs2011, merendaAugmentedRealityInterface2018}.

Recent advances in ubiquitous sensing systems have enabled precise capture of individual and group behaviors across diverse environments, including physiological sensing for understanding affective states and wearable devices for tracking interpersonal synchrony. For a detailed discussion on sensing individual and group behavior in ubiquitous systems, including applications in VR and MR, please refer to \autoref{sec:appx:sensing_behavior}. To our knowledge, no research has yet explored passive and automated group behavior characterization in MR with sensors in modern headsets, and this work aims to fill that gap.

While existing studies have significantly advanced our understanding of individual behavior in mixed physical-digital environments, there remains a critical gap in comprehending group dynamics within MR settings. \sysname addresses this gap by leveraging the sensory capabilities of MR headsets to investigate group behavior in collaborative tasks. Our framework uniquely explores passive and automated group behavior characterization in MR, filling a void in current research.

\subsection{Dataset for Group Behavior in Collaborative Immersive Environment}
Existing open-source datasets, while valuable, do not adequately capture group behavior in collaborative MR tasks. For instance, the Aria Synthetic Environments Dataset lacks human interactions, while the Aria Digital Twin and Everyday Activities Datasets are limited to single-user or two-person interactions~\cite{projectaria}. Other datasets like Stanford 2D-3D-Semantics and UCSD MR OpenRooms focus on static environments or limited user interactions~\cite{armeni2017joint,li2021openrooms}. Even the recent Egocentric Concurrent Conversations Dataset~\cite{ryan2023egocentric}, despite including 50 participants, only captures social conversations without collaborative virtual tasks. Given that small-group behavior studies typically require 3-4 members per group, these existing datasets fall short of our research needs. To address this gap, we designed a framework for collecting data specifically tailored to study group collaboration in MR environments. As a key contribution of this paper, we conducted a user study\footnote{Our user study follows an approved IRB protocol.} creating a new dataset with 48 participants forming 12 groups, each collaborating on the same task independently. This dataset enables us to analyze group behavior in MR collaborative setups more effectively. Details of this user study are presented in Section~\ref{sec:user}.

\subsection{Sociometry and Group Behavior Analysis}\label{sec:sociometry}
Social Network Analysis (SNA), originating from sociometry~\cite{moreno1941foundations}, offers a powerful approach to studying group dynamics. Unlike traditional research focusing on individual attributes, SNA emphasizes relationships connecting social actors~\cite{yang2016social-historySNA}. This network approach provides a systematic method for analyzing social interactions through sociograms, which visualize group structures and relationships~\cite{moreno1941foundations, wasserman1994social-faust-1994}. Sociometry allows researchers to identify social structures and roles within a group, such as leaders, followers, and isolates, by analyzing the frequency and quality of interactions~\cite{hackman2010group-main-paper}. Sociograms have proven invaluable in diverse fields, including nursing~\cite{drahota2008sociogram-as-an-analytical-tool, baiardiUsingSociogramsEnhance2015, royseUsingSociogramCharacterize2020}, collaboration~\cite{appletonUseSociogramsExplore2013}, social network data entry~\cite{hoganEvaluatingPaperScreenTranslation2016}, organizational studies~\cite{wasserman1994social-faust-1994}, and gaming~\cite{herkenrath_geo-sociograms_2014}. In the context of virtual environments, SNA concepts have been applied to understand social interactions. Bailenson et al.~\cite{bailenson2004transformed} used virtual gaze to model user networks and detect leaders in collaborative virtual environments. SNA has also been shown to be effective in smaller, intimate settings such as classrooms~\cite{martinezCombiningQualitativeEvaluation2003}, team sports~\cite{lusherApplicationSocialNetwork2010}, and small group interactions~\cite{sauerMeetingsNetworksApplying2013a}. Yang et al.~\cite{yangImmersiveCollaborativeSensemaking2022} enhanced VR data analytics by visualizing social networks based on gestures and location proximity. \sysname builds upon this foundation, leveraging SNA to study group behavior in MR collaborative tasks. 

\begin{figure}[t]
  \includegraphics[width=\linewidth, trim={0.5cm 16cm 2.5cm 0},clip]{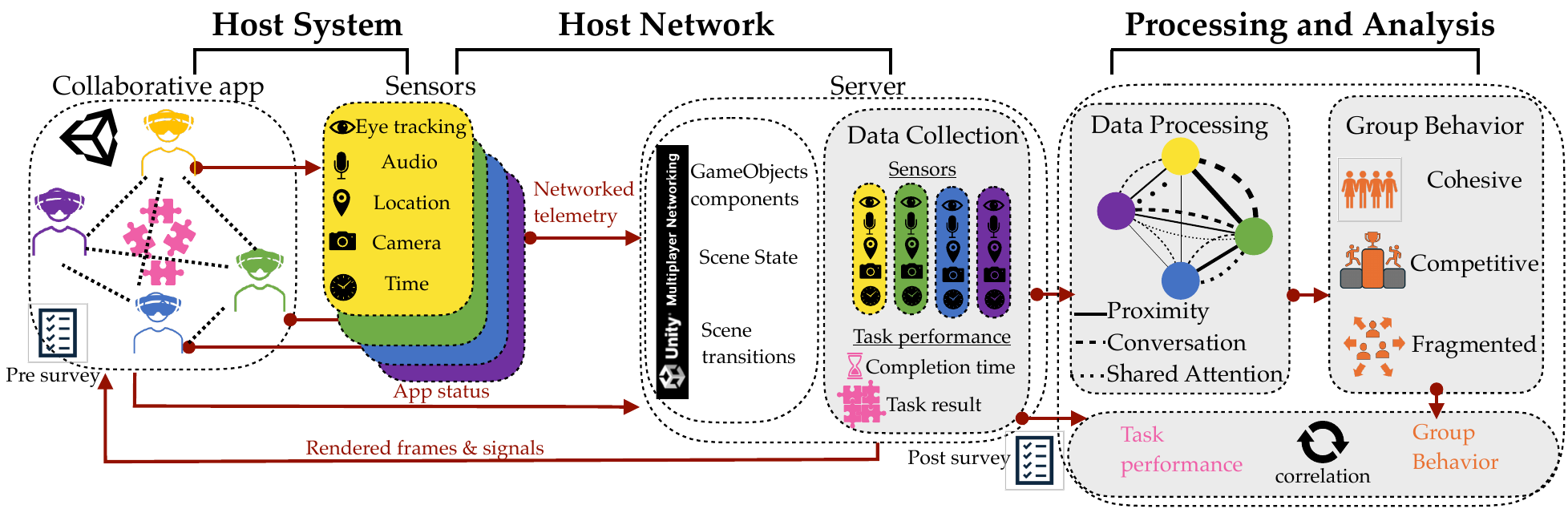}
  \vspace{-0.7cm}
  \caption{Multiple users collaborate using immersive MR applications. Their collaboration creates social dynamics reflected in sensor data captured by each MR device, which is communicated over a host network to manage the shared scene state and transitions. Pre- and Post-surveys are collected for each user. \sysname processing and analysis modules work offline to convert the raw sensor data into high-level proximity, conversation, and attention information to create a sociogram for analysis. Group behavior can be cohesive, competitive, and fragmented. Post-hoc analysis explores the correlation between the task performance (time to complete a task and accuracy/efficiency of the task) and the inferred group behavior. }
  \vspace{-0.1cm}
  \label{fig:framework}
\end{figure}

\section{Design and Implementation of \sysname Framework} \label{sec:framework}

We developed \sysname to study group behavior in collaborative MR tasks, combining headset sensors, software, and network analysis. This framework includes three modules: (1) detection, (2) processing, and (3) analysis, as described in~\autoref{fig:framework}. The sensing module collects data from headset sensors like microphones, location trackers, and eye trackers to capture group interaction signals such as conversation, shared attention, and proximity. The processing module converts this data into sociograms that map social interactions to reveal group behavior patterns. The analysis module uses social network analysis on the sociograms to interpret group dynamics, providing insights into cohesion, competition, and fragmentation. With these components, \sysname provides a passive and automated framework to understand group behavior in MR systems.

\subsection{Passive Sensing for Interpersonal Behavior in MR}\label{sec:sensingmodule}

This module focuses on using the rich sensory system and software capabilities available on an MR headset to passively collect sensory data, including eye tracking, audio, location, and time on the device. We use these raw sensor data to capture information related to group behavior. In particular, we are interested in observing information related to three types of interaction, namely conversation, shared attention, and proximity, as they have shown relevance to unraveling group behavior in various works in the literature, as detailed below.

\begin{enumerate}[leftmargin=0.6cm, topsep=0.1cm]
    \item[\textbf{(a)}] \emph{Conversation} is a key metric for studying small group behavior in psychology, anthropology, and observational studies~\cite{bales1950interaction,levine1990progress, brown2019group,pentland2012new}. Recent research using ubiquitous sensors continues to emphasize its importance~\cite{samroseCoCoCollaborationCoach2018, leeFlowerPopFacilitatingCasual2018, zhang_teamsense_2018, voletIndividualContributionsStudentled2017}. Furthermore, conversation patterns vary across different media, including virtual reality and desktop environments~\cite{yangImmersiveCollaborativeSensemaking2022}. MR immersive experiences uniquely blend virtual and physical interactions, making speech and conversation metrics crucial for understanding group behavior. To this end, the sensing module leverages the audio data collected from the microphones embedded in the MR headsets to capture conversation metrics.

    \item[\textbf{(b)}] \emph{Shared attention,} 
    a cognitive process involving the ability of individuals to focus with a social partner on a shared object of interest for intentional or social purposes has been extensively studied in the literature~\cite{bakemanCoordinatingAttentionPeople1984, mooreJointAttentionIts1995}. Various work in the literature has consistently highlighted the critical role of joint attention in facilitating cooperation and social bonding~\cite{yuillHowGettingNoticed2014, wolfJointAttentionShared2016, dejaegherParticipatorySensemaking2007}. Given its pivotal role in collaborative behavior, the significance of joint attention in the context of computer-supported collaborative work has been a subject of considerable exploration~\cite{yiqiuzhouCharacterizingJointAttention2022, jingImpactSharingGaze2022b, vertegaalGAZEGroupwareSystem1999a}. Many studies have used gaze awareness for automatic joint attention detection. Building on this, our \sysname framework incorporates joint attention using eye-tracking data from MR headset sensors.

    \item[\textbf{(c)}] \emph{Proximity} 
     has been shown in studies to positively correlate with knowledge transfer, creation, teamwork quality, and collaboration in studies related to organizational behavior~\cite{vissersKnowledgeProximity2013, hoeglTeamMemberProximity2004, krautPatternsContactCommunication1988}. Specifically, research has revealed that during collaborative co-location scenarios, collaborative benefits are enhanced when participants are placed in close proximity rather than at a distance~\cite{hawkeyProximityFactorImpact2005, valacich1994physical}. In addition, studies have shown that individuals tend to be physically closer to each other when their relationships are more intimate~\cite{cristaniComputationalProxemicsInferring2011}. Therefore, interpersonal proximity is a significant factor in characterizing group behavior. Consequently, the \sysname framework monitors the proximity of group members by tracking the relative location of the MR headset of each participant within the collaborative MR setup. 

\end{enumerate}

\subsection{Data Processing for Modeling Pairwise Interactions With Sociograms}\label{sec:processmodule}
An overview of the second module of our \sysname framework can be seen in \autoref{sec:appx:data-proc}~\autoref{fig:data-processing-summary}. This module models the raw sensor data collected from the MR headset into a format that can be processed. Hence, our approach is modeling this data as pairwise interactions to uncover underlying patterns in group behavior.
In our sociograms, the nodes represent individual users in the collaborative MR task, while the edges represent pairwise interactions. Edges can be directed (one-way relationships) or undirected (mutual relationships), with weights indicating the intensity of the interaction based on frequency or duration. For each type of interaction, we explain (1) the challenges in capturing the required sensor modality and (2) the data processing pipeline from raw sensor data to the sociogram. This systematic approach offers a valuable understanding of how interactions in MR environments can be passively and automatically detected using commercial devices.

\subsubsection{Modeling Verbal Interaction with Sociograms.} 
We generated sociograms based on \emph{speaking duration}, a metric proven effective in analyzing group behavior~\cite{drahota2008sociogram-as-an-analytical-tool, baiardiUsingSociogramsEnhance2015, royseUsingSociogramCharacterize2020}. The primary challenge in conversation analysis is accurately identifying the target of a speaker's utterance. Previous research~\cite{jiaAudioVisualConversationalGraph2024} shows that even with labeled data and specific instructions, detecting directed conversation accurately is difficult. Our pilot study (\autoref{app:pilot}) reinforced this, as participants often spoke to the entire group rather than specific individuals. This aligns with existing research suggesting that participants in MR focus more on virtual objects than direct social interaction~\cite{schnier_collaboration_2011}. To address these limitations, we assumed participants spoke to the entire group when initiating a conversation. This process results in a conversation sociogram, a directed graph in which nodes represent participants, edges denote conversation direction, and edge weights reflect interaction duration in seconds. For details on the data processing pipeline, including conversation sociogram generation, audio processing, and speaker identification, see Appendix~\ref{sec:appx:conv=data-proc}.

\subsubsection{Modeling Shared Attention by Gaze-Based Interaction with Sociograms.} 

Building upon prior research in shared attention and gaze awareness~\cite{yiqiuzhouCharacterizingJointAttention2022, jingImpactSharingGaze2022b, vertegaalGAZEGroupwareSystem1999a}, \sysname framework introduces an MR-specific shared attention sociogram. This model leverages the eye-tracking data from MR headsets to analyze how gaze alignment on virtual objects influences collaborative group behavior and task effectiveness in immersive contexts.

Developing a gaze-based interaction sociogram presents several challenges. Research suggests that eye gaze behaviors differ between AR/MR head-mounted devices and traditional face-to-face interactions~\cite{prytzImportanceEyecontactCollaboration2010}, with users in AR/MR often focusing more on virtual artifacts than direct eye contact. To address these challenges, we focused on shared attention on virtual objects as the primary indicator of gaze interaction. Another challenge is the determining appropriate thresholds for gaze duration on a virtual object, as brief glances may not signify meaningful shared interaction. To filter out such fleeting eye movements, we considered gaze durations exceeding 13 milliseconds, aligning with research on visual perception~\cite{potterDetectingMeaningRSVP2014}. This process results in an undirected gaze-based interaction sociogram where nodes represent participants, edges represent shared attention on virtual objects, and edge weights indicate the cumulative duration of shared gaze. For detailed the shared attention sociogram data processing pipeline, including eye-tracking, gaze logging, and overlap calculation, see Appendix~\ref{sec:appx:gaze-proc}.

\vspace{-0.1cm}
\subsubsection{Modeling Proximity-Based Spatial Interaction with Sociograms.} 

Previous research introduced geo-sociograms in gaming contexts, visualizing player movement and social dynamics through matrix-style plots~\cite{herkenrath_geo-sociograms_2014}. However, these methods don't align directly with traditional sociogram analysis. We propose a proximity sociogram that captures spatial relationships between participants in a collaborative MR environment, maintaining consistency with traditional sociograms. This approach allows us to explore how physical closeness influences group behavior, collaboration, and communication within the MR setting.

Accurate modeling of spatial interactions requires careful selection of proximity thresholds and sampling rates. We set the threshold at 1.5 feet, based on proxemic theory's definition of intimate space~\cite{rakel_chapter_2012}. A sampling rate of one second was chosen, considering the average duration of a gait cycle for normal walking speed~\cite{romero2023gaitguard}. This approach balances temporal resolution and computational efficiency, capturing transient interactions without overburdening the system. For details on the data processing pipeline for generating the proximity sociogram, including spatial data collection, synchronization, and distance calculation, please refer to \autoref{sec:appx:prox-proc}. This process results in an undirected graph where nodes represent participants, edges indicate proximity, and edge weights represent the duration of being within 1.5 feet of each other. This approach provides insights into spatial relationships, interaction patterns, and group cohesion, which will be further analyzed in \S\ref{sec:analysismodule}.

\section{Data Analysis Methodologies to Interpret and Characterize Group Behavior}\label{sec:analysismodule}

In the third module of our \sysname framework, we develop a tool to characterize group behavior using three distinct sociograms. To achieve group-level characterizations with sociograms of multiple interaction types,
we face multiple challenges, including (1) \textbf{CH1} heterogeneous data integration, (2) \textbf{CH2}  interpreting pairwise interactions for group-level insights, (3) \textbf{CH3} selecting metrics that depict group-level characteristics, and (4) \textbf{CH4} aggregating the low-level metrics to high-level insights that can translate the sociograms into collective group behavior labels. We assess group behavior by aggregating categorical scores from sociograms into \texttt{cohesive, competitive, or fragmented} labels~\cite{wasserman1994social-faust-1994}, where high cohesion and balanced interactions indicate cohesion, disconnection suggests fragmentation and centralized influence reflects competition, ensuring a weighted, data-driven understanding of group dynamics through sociometric analysis~\cite{watts1998collective-dynamics-cohesion, wellman1988social}.

\vspace{-0.1cm}
\subsection{Heterogeneous Data Integration} Sociograms for conversation, proximity, and shared attention capture various aspects of group behavior. Unifying them into a single group label is complex due to varying data representation, completeness, context relevance, and reliability.To address these, we use graph theory as the mathematical basis for analyzing sociograms~\cite{drahota2008sociogram-as-an-analytical-tool}, unifying various interaction analyses into a cohesive framework. As noted in \S\ref{sec:related}, SNA uses graphs to represent actors and relationships, with nodes as individuals and edges as interactions~\cite{pereira2007improving-social-network-analysis}. It applies graph theory to evaluate properties such as centrality, betweenness, and closeness~\cite{zorrilla2019sociograms-as-a-tool}, with measures being individual (local) or graph-level (global)~\cite{wasserman1994social-faust-1994}. Individual measures highlight key participants, while graph-level measures assess the structure and cohesion of collective behavior~\cite{fortunato2010community}. We interpret the sociograms, despite their heterogeneity, by applying consistent graph-based metrics to reveal underlying group behavior.

\subsection{Interpreting Pairwise Interactions} 

We examine collaboration through its social aspects, focusing on group-level interactions~\cite{dowson2003students-mcinerney}. Our priority is global measures~\cite{dillenbourg1999collaborative, johnson1989cooperation, network-paradigm-foster, wasserman1994social-faust-1994} like cohesion, influence, and connectivity, rather than individual factors like hierarchy~\cite{wellman1988social, baran1964distributed-category}. Translating pairwise interactions into cohesive group-level insights presents a significant challenge.  Each sociogram edge captures an interaction between two users, reflecting aspects such as communication intensity or shared attention. While valuable, these pairwise metrics don't inherently convey broader group structures or dynamics. For instance, strong connections between some members may coexist with weaker or missing interactions among others, potentially leading to group fragmentation. \textbf{Understanding group behavior requires assessing how these pairwise interactions combine to form subgroups, influence distributions, or group-wide cohesion.}
To quantify group interactions, we use weighted edges over unweighted ones, as they better capture the intensity and significance of relationships. This approach links group behavior to experience and task performance across interaction types~\cite{granovetter1973strength-weak-ties}.

\subsection{Metric Selection and Weighted Analysis} 

Selecting appropriate metrics is crucial for capturing group-level characteristics from our sociograms. We focused on metrics that incorporate interaction strength, avoiding those that don't support weighted properties (e.g., density, diameter, reciprocity). This approach allows for consistent analysis reflecting varied interaction intensities across different sociogram types.
\sysname computes metrics for a given sociogram graph $G$, represented by an adjacency matrix $A_{ij}$, where $A_{ij}$ contains the edge weight between nodes $i$ and $j$ (or $0$ if no edge exists).
We use weighted graph metrics to assess group structure and interaction patterns, helping categorize groups based on individual behavior within the group.
These metrics help identify group labels by analyzing member roles, interaction density, and structural dependencies. 
Methodologies and metric formulations are in~\autoref{alg:metric-computation}, with their role in group labeling summarized in~\autoref{tab:group_behavior_metrics}, both in Appendix~\ref{sec:appx:characterization}.
Next, we define the selected graph metrics and their role in identifying group behavior: 

\begin{enumerate}[leftmargin=0.6cm, topsep=0.1cm]
    \item[\textbf{(a)}] \emph{Eigenvector Centrality ($x_i$)} quantifies overall influence in the group by identifying key members based on both direct connections and the influence of their neighbors. It helps detect power dynamics and opinion leaders, with mean eigenvector centrality offering an average influence baseline, its variability indicating whether the influence is evenly distributed or concentrated.

    \item[\textbf{(b)}] \emph{Clustering Coefficient ($C_i$)} measures local group density by assessing how well-connected an individual's neighbors are. In a shared attention sociogram, high clustering coefficients can indicate subgroups working closely together or sharing a common focus. Variability in clustering coefficients reveals differences in local group density, highlighting areas of concentrated or sparse collaboration.

    \item[\textbf{(d)}] \emph{PageRank ($P(i))$} identifies members who drive discussions and maintain engagement within the group. It measures the influence based on the number and quality of connections, identifying those driving attention or discussions and contributing significantly. The damping factor ($d$) ensures realistic influence distribution.

    \item[\textbf{(e)}]\emph{Katz centrality ($\mu_{katz}$)} assesses cumulative influence, incorporating both direct and indirect connections. It highlights participants whose impact extends beyond immediate ties, revealing hidden influencers who shape group behavior. It considers all paths with a decay factor ($\alpha$), indicating cumulative influence from indirect connections.
    
    \item[\textbf{(f)}] \emph{Path Length Variability ($\sigma\_{PL}$)} distinguishes between central and peripheral members by analyzing the consistency of shortest paths. High variability suggests reliance on specific individuals for communication, indicating potential bottlenecks or uneven participation. 

    \item[\textbf{(g)}]\emph{Edge Connectivity ($\kappa'(G)$)} can represent the resilience of relationships or interactions within a sociogram. High connectivity suggests a strong, cohesive group structure that can withstand disruptions.
\end{enumerate}

\subsubsection{Multi-Step Heuristic Aggregation} 
Multi-step heuristic aggregation is necessary to transition from individual metric analysis to group behavior labels. This process poses challenges in capturing emergent group properties, as individual metrics may not fully represent collective dynamics. For example, high clustering variability could indicate group fragmentation despite strong pairwise interactions. Therefore, metrics must be interpreted collectively and in context to avoid misrepresentations.
\sysname processes metrics from each sociogram, pairing and aggregating them to identify high-level patterns such as cohesion, influence, and connectivity. 
To systematically analyze group behavior, we characterize it through these dimensions as they capture fundamental structural and functional properties of group interactions~\cite{wellman1988social,granovetter1973strength-weak-ties}.
Specifically, cohesion reflects how tightly members are connected, affecting stability and collaboration~\cite{moody2003structural-cohesion, watts1998collective-dynamics-cohesion}. Influence measures the distribution of decision-making, distinguishing between participatory and hierarchical structures~\cite{bonacich1987power-centrality}. Connectivity assesses the resilience of interaction networks, ensuring efficient communication and adaptability~\cite{holme2002edge}.

This integrated approach provides a holistic assessment of group dynamics beyond pairwise interactions, describing:
\begin{itemize}
    \item How well a group functions as a unit (cohesion).
    \item Who holds power in decision-making (influence).
    \item How robust the group is to disruptions (connectivity).
\end{itemize}

\subsubsection{Weighted Thresholds}
The selected thresholds for each metric are based on a combination of graph-theoretic principles, empirical observations, and insights into network dynamics. Each threshold is chosen to reflect meaningful distinctions in group behavior while maintaining robustness across different contexts. The detailed justification of the chosen characterizations and the metric threshold is presented in Appendix~\ref{sec:appx:charaterization} and~\ref{sec:appx:threshold-selection}, respectively; for brevity, we present here the finalized thresholds we used in our study as follows:

\begin{itemize}
    \item Cohesion is measured by clustering variability ($\sigma_C$) and eigenvector variability ($\sigma_{x_i}$): high if \( \sigma_C < 0.05 \) and \( \sigma_{x_i} < 0.08 \), moderate if \( \sigma_C < 0.02 \) and \( \sigma_{x_i} < 0.04 \), otherwise low. 

    \item Influence uses mean eigenvector centrality ($\mu_{x_i}$), PageRank variability ($\sigma_{P}$), and mean Katz centrality ($\mu_{katz}$): high if \( \mu_{x_i} > 0.47 \), \( \sigma_{P} < 0.02 \), \( \mu_{katz} > 0.465 \); moderate if \( \mu_{x_i} > 0.49 \) and \( \sigma_{P} < 0.065 \), otherwise low.

    \item Connectivity is evaluated by edge connectivity ($\kappa'(G)$) and path length variability ($\sigma_{PL}$): resilient if \( \kappa'(G) > 2 \) and \( \sigma_{PL} < 24 \), moderate if \( \kappa'(G) > 1 \) and \( \sigma_{PL} < 50 \), otherwise fragile.
\end{itemize}

\begin{algorithm}[t]
\small
\caption{\texttt{GroupScorer}: Calculate Scores and Assign Group Label. 
}
\label{alg:final-group-label}
\KwIn{$group\_characteristics$ for conversation, proximity, and shared attention; weights configurations $weights$}
\KwOut{$group\_label$ as $category \in \{cohesive, competitive, fragmented\}$}
Initialize $score[cohesive]$, $score[fragmented]$, $score[competitive]$ to 0\;
\ForEach{$(graph\_type, characteristic)$ in \{(conv, c\_character), (prox, p\_character), (att, att\_character)\}}{
    \ForEach{$(characteristic, value)$ in $characteristics$}{
        \If{$value \in \{high, distributed, tight-knit\}$}{
            $score[cohesive] \gets score[cohesive] + weights[graph\_type]$\;
        }
        \ElseIf{$value \in \{low, fragile, loose-knit\}$}{
            $score[fragmented] \gets score[fragmented] + weights[graph\_type]$\;
        }
        \ElseIf{$value \in \{resilient, moderate, centralized\}$}{
            $score[competitive] \gets score[competitive] + weights[graph\_type]$\;
        }
    }
}
\Return $group\_label \gets \underset{category}{\arg\max} \, score(category)$
\end{algorithm}

\subsection{Final Group label Generation}
The final step in our process involves assessing group behavior by aggregating scores derived from categorical attributes observed in each sociogram. We focus on three overarching group behavior categories: \texttt{cohesive}, \texttt{competitive}, and \texttt{fragmented}, which capture key dimensions of social dynamics in collaborative environments~\cite{wasserman1994social-faust-1994}. These categories reflect distinct group characteristics: \texttt{cohesive} groups demonstrate balanced and well-integrated structures with evenly shared interactions, indicating strong connectivity and mutual support; \texttt{fragmented} groups show signs of disconnection, suggesting potential issues with integration and communication that may hinder overall performance; \texttt{competitive} groups present a complex dual nature, potentially exhibiting hierarchical structures that, while reducing cohesion, can be advantageous in task-focused settings by enabling efficient decision-making and highlighting key influencers.

The label assignment process, summarized in \autoref{alg:final-group-label}, involves iterating through group characteristics derived from each type of sociogram and adjusting scores for the broader group categories. High cohesion and tight-knit clustering indicate \texttt{cohesive} group behavior, while low cohesion or loose-knit clustering suggests \texttt{fragmentation}. Attributes such as high influence or a centralized structure contribute to \texttt{competitive} behavior. We compute aggregate group scores using specific rules: high, tight-knit, or distributed labels boost the \texttt{cohesive} score; low, fragile, or loose-knit increase the \texttt{fragmented} score; and resilient, moderate, or centralized labels raise the \texttt{competitive} score, indicating robustness or balance. The highest-scoring group behavior becomes the assigned label for that configuration.

Importantly, configurations can be empirically weighted to highlight different interactions, revealing how various interaction types shape group behavior measurement. \sysname's comprehensive approach to profiling group behavior through sociometric data and graph analysis enables each interaction type to contribute to understanding group dynamics. By prioritizing weighted metrics, we capture the quality and significance of interactions, enabling an in-depth analysis of group behavior that goes beyond simple pairwise interactions and provides a holistic view of group dynamics in collaborative environments.

\begin{table}[!t]
  \centering
  \small
  \caption{Participant Demographics. The key for frequency: never/almost never; rarely (<2times); occasionally (a few times); frequently in the past; frequently (> 2times/month). 
  }
  \vspace{-0.4cm}
  \label{table:demographics}
  \begin{tabular}{|l|c|}
  \toprule \hline
              \makecell[l]{\textbf{Demographics}} & \makecell[c]{\textbf{Number of Participants}}  \\ \hline  \hline
      
          \makecell[l]{Gender} & \makecell[c]{36 Male, 12 Female}  \\\hline
          
          \makecell[l]{Frequency of AR Experience} & \makecell[c]{22 Never Used, 15 Rarely, 6 Occasionally, 3 Frequently, 2 Frequently in the past}  \\\hline
          
          \makecell[l]{Frequency of VR Experience} & \makecell[c]{21 Never Used, 13 Rarely, 7 Occasionally, 6 Frequently, 1 Frequently in the past}  \\\hline
          
          \makecell[l]{Frequency of Gaming} & \makecell[c]{4 Never Used, 10 Rarely, 15 Occasionally, 16 Frequently, 3 Frequently in the past}  \\\hline
          
          \makecell[l]{Familiarity to Other Members} & \makecell[c]{25 No Members, 15 One Member, 6 Two Members, 2 Three Members}  \\\hline
          \bottomrule
  \end{tabular}
  \end{table}

  \section{User Study}\label{sec:user}
  Group behavior in small groups has been studied in fields such as psychology, anthropology, and sociology but has not been explored in MR~\cite{bales1950interaction,tuckman1977stages, turner2017ritual,blau2017exchange}. Other works, such as TeamSense~\cite{zhang_teamsense_2018}, which focused on cohesion sensing during a simulated long-term Mars mission, highlight the potential of sensor-based group analysis. However, longitudinal setups are rare in MR scenarios due to HMD fatigue and battery limitations. To this end, we conduct a human-subject study to observe group behavior while users interact in an MR-based collaborative environment, validating and evaluating \sysname. Before conducting our main study, we conducted the pilot study (\autoref{app:pilot}) to test our software and refine the placements of the virtual objects. 
  
  \subsection{Participants}
  We recruited $60$ participants; data from 12 participants were discarded due to technical issues, resulting in $48$ participants being included in the analysis. Participants were allowed to form their own groups, or, if preferred, the research team randomly assigned them to a group. In total, 12 groups completed the study. In similar research on collaborative behavior involving small groups, the group is typically defined as having three or more members~\cite{university_82_nodate}; in our study, we formed a group with 4 members. Prior studies have revealed patterns of conversation, interaction, and coordination within groups of four in both desktop and virtual reality environments~\cite{yangImmersiveCollaborativeSensemaking2022}. Most prior research has focused on dyads or triads~\cite{slaterSmallGroupBehaviorVirtual2000a, bailensonGazeTaskPerformance2002, basdoganExperimentalStudyRole2000, podkosovaCopresenceProxemicsShared2018, numanExploringUserBehaviour2022, abdullahVideoconferenceEmbodiedVR2021}. By adopting a group size of four participants, we aimed to create a richer collaborative environment, capturing the complexities of group behavior not as evident in smaller groups.
  Participants' ages ranged from 21 to 42, with a mean age of 24. A summary of the participant demographics can be seen in~\autoref{table:demographics}. 
  All participants provided informed verbal consent.
  Each participant had normal or corrected-to-normal vision. The institution's ethics committee approved the study.

  \subsection{Materials}
  To capture sensor data on user interactions, we used the Meta Quest Pro~\cite{MetaQuestProa}. The integrated sensors enabled audio recording, eye-tracking, six-degrees-of-freedom (6 DoF) simultaneous localization and mapping (SLAM) tracking, and mixed reality capabilities. The collaborative application is developed using Unity with Meta XR packages~\cite{metaXRpackages}. Unity and Meta XR API's are used in custom scripts to record the audio, location, gaze-object intersections, and user interactions with virtual objects. For precise manipulation of virtual objects, we use the controller integrated with the Meta Quest Pro~\cite{MetaQuestProa}. The interaction recording and tracking is validated by others~\cite{published_are_2022}.
  Participants were invited to a shared lab room with a designated $10\texttt{ft} \times 5\texttt{ft}$ space cleared of materials to minimize distractions. They were informed they could move freely within this area during the task.
  
  \subsection{Experimental Task}\label{sec:taskdesign}
  This section summarizes the cue, interaction, and feedback of our collaborative image sorting task. Each group of four participants completed one image-sorting task using the same images and categories. Participants were instructed to work together to reach a consensus on the grouping of each image.
  
  \subsubsection{Primary Task} Participants were tasked with sorting 28 carefully selected images from the Open Affective (OASIS) dataset~\cite{kurdi_introducing_2017}. The OASIS dataset, comprising 900 validated images by over 822 participants for pleasantness and arousal ratings. The images were selected to represent a range of emotions while excluding potentially distressing content such as violence or graphic imagery. Participants were tasked with sorting the selected images into one of six emotion categories randomly chosen from Russell's circumplex model of affect~\cite{russell_circumplex_1980}. These categories included angry, bored, relaxed, tense, pleased, and frustrated. 
  Image sorting tasks have been shown to foster decision-making, communication, and social coordination by building shared mental models and group alignment~\cite{rorissaFreeSortingImages2004, bjerreCardSortingCollaborative2015, berlinFeedbackCooperationExperiment2024}. 
  We focus on this collaborative task marked by asynchronous, flexible participation, where participants can contribute and modify inputs independently. This repeated image-sorting task, involving open-ended discussions on the emotions elicited by each image, allows us to observe a dynamic, iterative collaborative process among the group, where participants achieve a shared goal through incremental steps and consensus. Each group of four participants completed one image-sorting task on the same 28 images and categories. Participants were instructed to work together as a group to reach a consensus on the label of each image, with no time limit for completing the task, allowing participants to engage in deliberate discussion and negotiation. The labels are not placed in a fixed position, allowing participants to organize and use the room space however they want. The task ends once they all report that they agree with the image groupings.
  
  \subsubsection{Virtual Scene and Cues} As shown in ~\autoref{fig:showcase} (left, headset's first-person perspective), all 28 virtual images are scattered around the room, and all six emotion categories labels (gray virtual plates) are pasted a little higher than where the images are scattered. Participants can view these images and labels at all times via their headsets. Participants determine the cue to start interactions by selecting images, allowing open collaboration without predefined structure or researcher direction.
  This lack of structure in cues is by design, as our goal is to observe open and free collaborations without participants taking turns or directed by the flow of the virtual scene designed by the researcher. The participants are assumed to take the cue for virtual interaction from the other three group members as shown in ~\autoref{fig:showcase} (right), where all four participants are physically close, probably examining the same image and discussing the final label. 
  
  \begin{wrapfigure}{r}{0.24\textwidth}
  \vspace{-0.3cm}
  \centering
    \includegraphics[width=\linewidth]{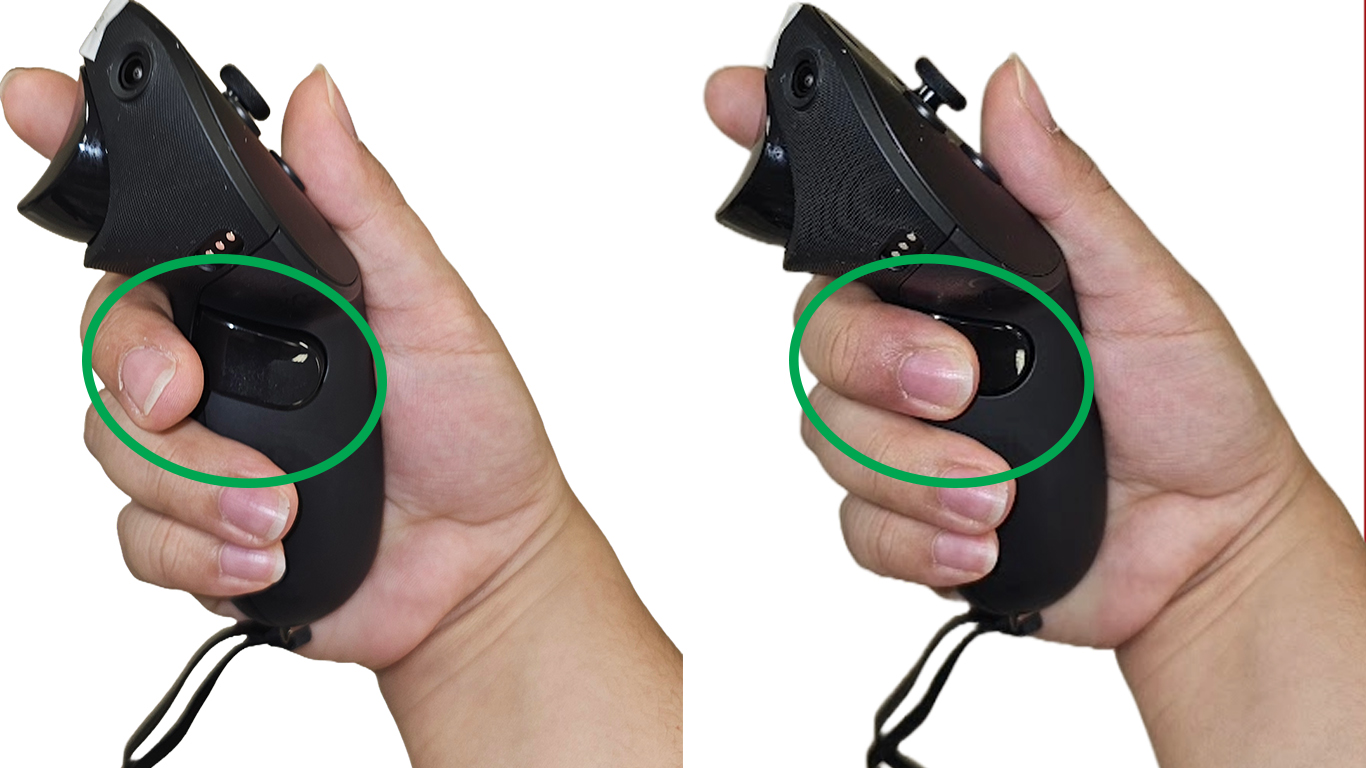}
    \vspace{-0.7cm}
    \caption{Controller gesture for image and label grabbing.}
    \vspace{-0.5cm}
    \label{fig:task-interaction-demo}
  \end{wrapfigure}
  \subsubsection{Interaction and Feedback} To sort an image, the participants were asked to move the virtual image near the virtual label physically. Once the image is pasted close to the label, the image is recorded as sorted. To move an image to the corresponding label, participants used a point-and-drag near-interaction motion with the grip buttons on their left or right controllers. They pointed the controller at an image, pressed and held the grip button, and moved the image by guiding the controller. The image followed the controller's pointer until the grip button was released, locking the image in place at its final position. Participants could only grab objects within reach and were instructed to release the grip button to secure an image in place, as demonstrated in ~\autoref{fig:task-interaction-demo}. This interaction closely mimics the physical action of picking up and placing objects. Once the image was positioned next to the label, it remained stationary when the grip button was released, providing feedback that the image was placed as intended. Only one participant could move each object at a time, but all were allowed to hold different image objects at the same time.

  \subsection{Measurements}\label{sec:measure}
  This section outlines the measures we used to capture group behavior in the image sorting task, categorized into sensor level, task-related, performance, and subjective measures. At the sensor level, we collected data from the headset using custom scripts. We recorded audio signals from the microphone, $x,y$ positions for location, and eye-tracking data to capture the data for conversation, proximity, and shared attention sociogram as the described 
  in~\S\ref{sec:sensingmodule}.

  On the task level, we recorded the various virtual object interactions, such as the number of virtual image interactions per participant in a group, to count if a participant grabs a virtual image and then releases it. Throughout the task, we capture the number of label changes per group to count the number of times a particular image changes its label. We also captured distinct groupings for each image per group to count the distinct labels for each image. For instance, if Participant A moved an image to label X, participant B moved it to label Y, and Participant A moved it back to X, the image would have three label changes and two distinct groupings. We also collected high-level performance metrics such as completion time. We measured completion time as the time elapsed from when the first image was grabbed to when the last image was placed, indicating the group's overall time completing the task. 
  
  We also collected high-level performance metrics such as subjective accuracy and completion time. We measured completion time as the time elapsed from when the first image was grabbed to when the last image was placed, indicating the group's overall time completing the task.
  We measure accuracy as the proportion of correctly categorized images into the corresponding emotion terms. The ground truth emotion term for each image was determined by matching the distribution of each OASIS image in a valence-arousal matrix to its equivalent coordinate in Russell's affect model, which organizes emotion terms by valence-arousal~\cite{russell_circumplex_1980}. The labels of these images are highly subjective, and hence, the accuracy can be below 60\% as reported in another study that used the same dataset~\cite{yangImmersiveCollaborativeSensemaking2022}. For us, this subjective labeling was part of our design and essential to encourage collaboration. Asking participants to categorize images based on content, such as living or non-living, would not prompt meaningful discussion, as such decisions are trivial. Including subjective accuracy in our work is crucial for understanding how well users align with shared emotional conventions during collaboration. It provides insight into the effectiveness of group consensus-building, the extent of divergence in interpretations, and how interaction dynamics influence collective decision-making. This measure helps us evaluate the role of discussion and negotiation in forming shared mental models, a core aspect of our study. We discuss more on this subjective accuracy in Appendix~\ref{sec:appx:accuracy}.
  
  Finally, we collect several subjective measures through post-exposure surveys after the image sorting task, such as presence (subjective feeling of being present in a virtual environment) with the IPQ~\cite{schubert_experience_2001} and PQ~\cite{witmer_measuring_1998} questionnaires, cognitive load through NASA-Task Load Index (NASA-TLX)~\cite{hart_nasa-task_2006}\footnote{We are aware of the criticism surrounding the use of NASA-TLX. We applied it as intended to measure ``perceived" cognitive workload, rather than actual mental load~\cite{mckendrick2018deeper}.}, and perception of the group behavior through a custom-designed questionnaire. We measure presence with the IPQ~\cite{schubert_experience_2001} and PQ~\cite{witmer_measuring_1998}. The PQ evaluates factors such as the possibility to act and examine, realism, self-evaluation, and interface quality. The IPQ measures factors such as spatial and general presence, realism, and involvement. The presence scores are derived from 33 items (14 IPQ and 19 PQ, the cognitive load score is derived from 5 items from NASA-TLX, and group behavior from 5 items from our custom-designed survey on a 7-point scale. We developed a custom questionnaire to measure participants' perspectives of how their group interacted, as shown in~\autoref{table:custom-questionnaire}. The proposed group behavior characterization questionnaire is designed to assess key aspects such as contribution awareness, joint attention, proximity impact, conversational support, and overall group collaboration. Each question is informed by established research to ensure relevance to our study's context. The purpose of this questionnaire is to capture subjective insights into how participants perceived the interactions measured by different modules in \sysname. This allowed us to validate our group behavior labels by comparing data from objective sensor data in sociograms with participants' subjective experiences.

  \begin{figure}[t]
    \centering
    \includegraphics[width=0.8\linewidth]{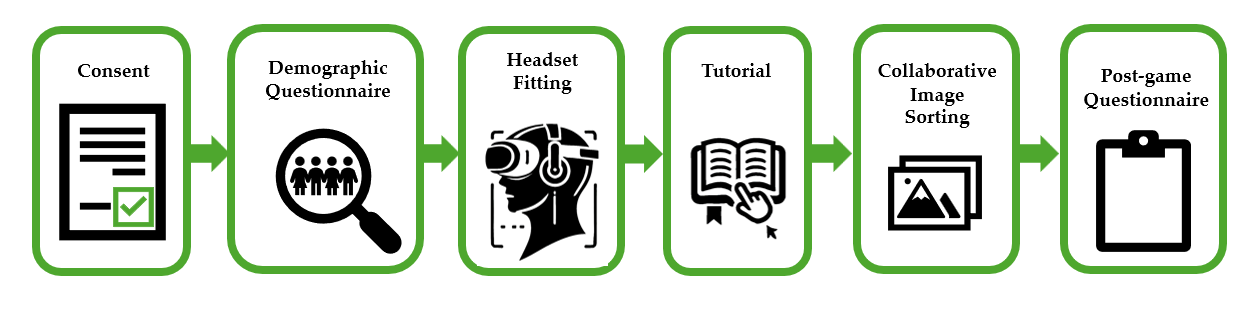}\vspace{-0.6cm}
    \caption{User study procedure: Participants provided consent and completed a demographics survey followed by headset calibration, a tutorial, the main collaborative image sorting task, and a post-exposure questionnaire.}\vspace{-0.1cm}
    \label{fig:user-study}
  \end{figure}
  \subsection{Study Procedure}
  The procedure for the study involved several steps, as shown in~\autoref{fig:user-study}. Upon arrival at the laboratory, participants were given an IRB-approved study information sheet containing detailed information about the study procedures, data collection, and privacy measures. The administering researcher also verbally briefed them on this information. They were also verbally instructed on the headset interactions and controller gestures needed to complete those interactions required during the tasks, as well as how to perform them. The briefing included details about the visual stimuli used in the experiment, such as their color, shape, duration, cues, and feedback mechanism for the image sorting task. Participants were given ample time to consider their participation in the study and were asked for their verbal consent. Following the briefing, participants were asked to fill out a demographic questionnaire, with questions including their gender, age, familiarity with technology, and with other group members.
  
  MR headsets were then distributed to the participants, and they were instructed to calibrate the focus and fit of the headset for maximum comfort. Before the main task, participants completed a tutorial application with two images and two categories not included in the main task to prevent learning effects. This tutorial task aimed to familiarize them with the task interactions in terms of gestures and get comfortable with using the point-and-drag interaction from the controller. 
  Following this, the group proceeded with the main image sorting task, where they were tasked to sort $28$ images into six different categories. They were informed that there was no time limit for the task and that the main requirement was for them to reach a consensus on the image sorting for the task to end. To encourage a more natural and unconstrained group behavior, participants were not informed that they were being timed or evaluated on their accuracy.
  Upon task completion, a post-task questionnaire assessing their cognitive load, presence, and perspective of what they think about how the collaboration went with the other group members in the task condition.
  The time it took for each group to complete the task varied; however, in general, the total duration of the session, including consent, briefing, training, headset calibration, experiment, and questionnaires, took less than an hour.

\begin{figure}[t]
  \centering
  \begin{subfigure}[b]{\linewidth}
    \centering
    \includegraphics[height=3.2cm]{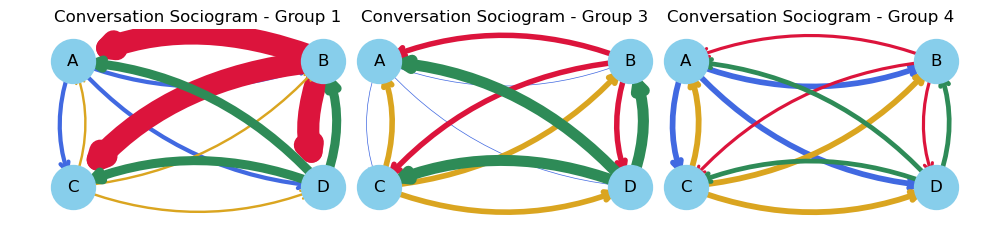}
  \caption{}
    \label{fig:subset-conversation-sociogram}
  \end{subfigure}
  \begin{subfigure}[b]{\linewidth}
    \centering
    \includegraphics[height=3cm]{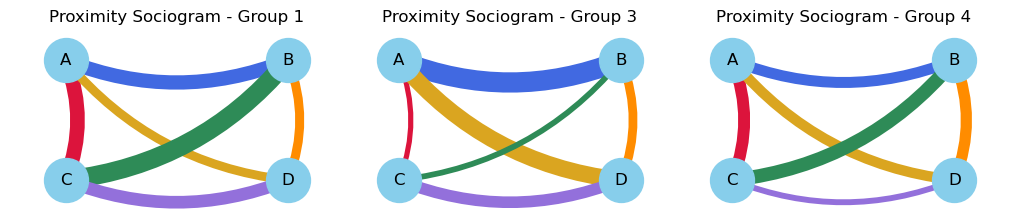}
  \caption{}
    \label{fig:subset-proximity-sociogram}
  \end{subfigure}
  \begin{subfigure}[b]{\linewidth}
    \centering
    \includegraphics[height=3cm]{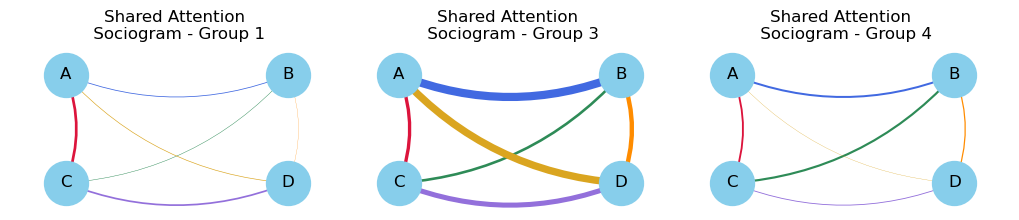}
  \caption{}
    \label{fig:subset-sharedatt-sociogram}
  \end{subfigure}
  \vspace{-0.7cm}
  \caption{Sociograms representing group interactions: (a) conversation, (b) proximity, and (c) shared attention. 
  In the conversation and proximity sociograms, the edge thickness reflects the total duration of respective interactions; the shared attention sociogram indicates the cumulative shared gaze on virtual objects.}
  \label{fig:combined-graphs}
\end{figure}

\section{Results}
\label{sec:results}
This section presents a comprehensive evaluation of \sysname in four categories. (1) Group and individual interaction analysis, offering qualitative insights into how groups and individual participants engage during tasks (\S\ref{sec:intrection-analysis}). (2) Group behavior profiling quantitatively examines the structural dynamics of group behavior depicted by sociograms across different modalities (\S\ref{sec:group-profiling}). (3) Statistical analysis investigates the significance of various task and behavior metrics to highlight differences and relationships that affect performance (\S\ref{sec:stat-eval}). (4) Inferential analysis, including mediation analysis, explains how group behavior influences task outcomes through indirect relationships between different tasks, performance, and experience elements (\S\ref{sec:mediation-analysis}). We are using data from 12 groups for our analysis and discarded the data from 4 groups due to technical issues.

\subsection{Group Interaction and Performance Analysis}
\label{sec:intrection-analysis}

\subsubsection{Sociogram Observations and Spatial Interactions}
\label{sec:sociogram-interactions}
This section analyzes group interactions through sociograms illustrating conversation, proximity, and shared attention, supplemented by participant movement, spatial use, and individual interaction counts.
\autoref{fig:combined-graphs} shows sample sociograms for (a) conversation, (b) proximity, and (c) shared attention, with full sociograms for all $12$ groups in Appendix~\ref{appendix:sociograms}.

Conversation sociograms reveal distinct speaking patterns: Group 1 displays mixed engagement, with Participants B and D leading while A and B contributed less, reflecting fragmented behavior. Group 3 shows Participant D's dominance, indicating uneven participation and potential hierarchical dynamics, often linked to competitive groups. Group 4 shows balanced speaking times, suggesting collaboration. 
These patterns suggest that evenly distributed speaking times (Group 4) imply stronger collaboration, whereas dominance by a specific member (Group 1) signals uneven participation, potentially impacting group decisions.
The proximity sociograms indicate different levels of participants' physical closeness. 
Group~1 exhibits strong proximity among participants A, B and C, suggesting possible subgroup formation. Group~4 shows balanced but less intense proximity, suggesting even engagement without clear subgroups. Group 3 displays notable proximity between Participants A and B and Participants C and D, indicating Participant A's key role in interactions. 
These observations suggest that evenly distributed proximity may correlate with higher collaboration, while concentrated proximity suggests subgroup behavior.
Shared attention graphs illustrate participants' focus on shared virtual objects. Groups 1 and 4 show minimal but evenly distributed shared attention, which can imply either fragmentation or cohesion. Group 4 indicates strong shared attention between sub-pairs (A-B, A-C, C-D), suggesting collaborative subgroups. Group 11 exhibits strong shared attention across all members, pointing to a coordinated effort. These patterns reflect varying levels of joint engagement, with interconnected attention signaling cohesion.

\begin{figure}
  \centering
  \includegraphics[width=\linewidth]{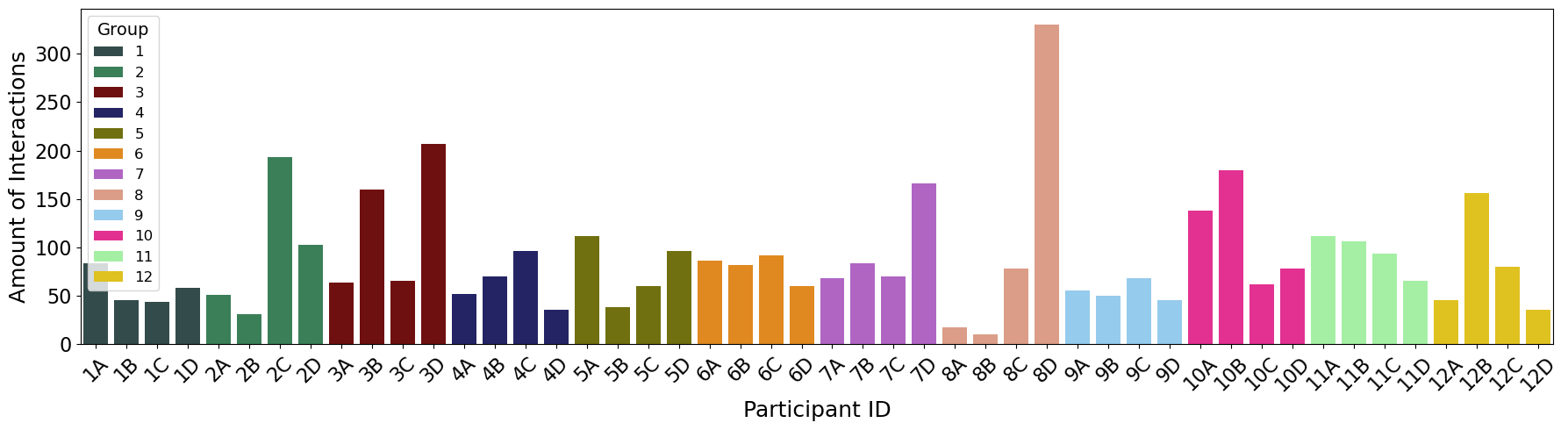}
  \vspace{-0.8cm}
  \caption{Number of image interactions per participant across all groups. }\label{fig:image-interactions}
  \vspace{-0.2cm}
\end{figure}
\begin{figure}
    \centering
    \begin{tabular}{cc}
    \includegraphics[width=0.68\linewidth]{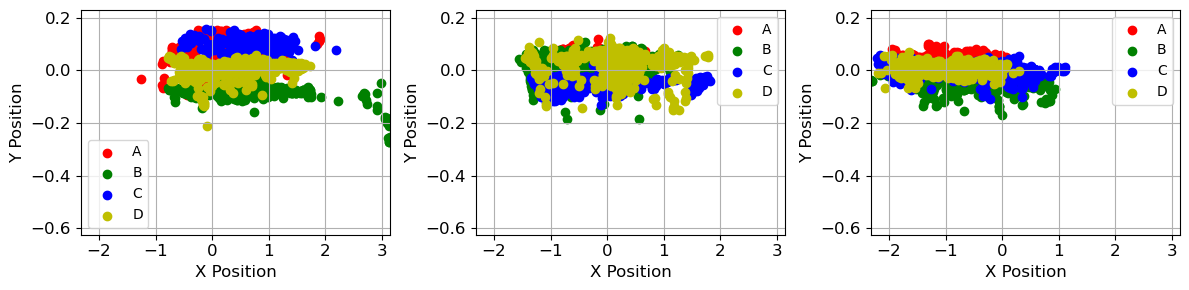}    &   \includegraphics[width=0.21\linewidth]{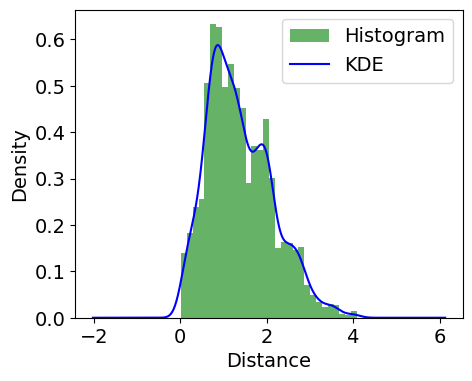}\\
    \quad\quad\quad (a) \quad\quad\quad\quad\quad\quad\quad\quad\quad (b) \quad\quad\quad\quad\quad\quad\quad\quad\quad (c) \quad\quad & (d)
    \end{tabular}
    \vspace{-0.3cm}
    \caption{Location heatmaps for the representative groups: (a) group 1, (b) group 3, (c) group 4, with the X and Y positions measured in meters. (d) shows the distance distribution between players across all groups. }
    \vspace{-0.1cm}
    \label{fig:location-heatmap}
\end{figure}

\autoref{fig:image-interactions} highlights significant variability in participant interactions across groups, ranging from $10$ to $113$. For instance, participants 8D and 2C show higher interaction counts, indicating dominant roles or intensive involvement, while 8B and 1D have lower counts, suggesting less participation. Group 2, with an average interaction count of $94.50$ and a high standard deviation of $72.34$, shows fragmented behavior due to uneven participation. In contrast, Group 11, with a similar mean but a lower standard deviation of $20.42$, indicates more balanced participation and greater cohesiveness. These interaction trends align with the sociograms, which reveal differences in conversation and shared attention among groups.
\autoref{fig:location-heatmap}a to~\autoref{fig:location-heatmap}c show spatial movement heatmaps for representative groups. Group 1 exhibits concentrated space use with participant clustering, Group 3 shows more dispersed movement, suggesting varied collaboration approaches, and Group 4 has a moderate spread, indicating a mix of individual focus and shared interaction.  
The distance distribution in~\autoref{fig:location-heatmap}d indicates most interactions occurred within close range, with density decreasing as distance increased. 

\begin{table}
    \centering
    \scriptsize
    \caption{Group-Level Task Metrics Summary.}
      \vspace{-0.4cm}
    \label{table:group-level-task-metrics}
     \resizebox{\textwidth}{!}{
    \begin{tabular}{|c||c|c|c|c||c|c||c|c||}
    \toprule
    \hline
   \textbf{\makecell[c]{Group \\ID}} & \textbf{\makecell[c]{Num of\\ Images \\Grabbed}} & 
   \textbf{\makecell[c]{Total Num \\ of Image \\Grabbing}}& 
   \textbf{\makecell[c]{Num of \\Image \\Labels \\Overridden}} & 
   \textbf{\makecell[c]{Total \\Images \\Looked \\At}} &
   \textbf{\makecell[c]{Completion \\Time \\ (seconds)}} & 
   \textbf{\makecell[c]{Accuracy (\%)}} &
   \textbf{\makecell[c]{Num of \\Label \\Changes}} \\ 
   \hline
   \hline
        1 & 50.0 & 232.0 & 22.0 & 109.0 & 415.54 & 67.86 \% & 56.0 \\  \hline
        2 & 52.0 & 378.0 & 24.0 & 112.0 & 620.59 & 57.14 \% & 72.0 \\ \hline
        3 & 71.0 & 497.0 & 43.0 & 111.0 & 676.78 & 53.57 \% & 88.0 \\ \hline
        4 & 51.0 & 254.0  & 23.0 & 110.0 & 513.09 & 42.86 \% & 55.0 \\ \hline
        5 & 54.0 & 306.0  & 26.0 & 110.0 & 209.26 & 50.00 \% & 61.0 \\ \hline
        6 & 60.0 & 320.0  & 32.0 & 112.0 & 622.31 & 60.71 \% & 71.0 \\ \hline
        7 & 71.0 & 388.0  & 44.0 & 111.0 & 562.77 & 44.44 \% & 101.0 \\ \hline
        8 & 58.0 & 436.0  & 30.0 & 112.0 & 994.05 & 17.86 \% & 59.0 \\ \hline
        9 & 50.0 & 220.0  & 22.0 & 112.0 & 430.89 & 60.71 \%  & 52.0 \\ \hline
        10 & 70.0 & 458.0  & 42.0 & 112.0 & 652.17 & 50.00 \%  & 84.0 \\ \hline
        11 & 72.0 & 378.0  & 44.0 & 112.0 & 534.00 & 42.86 \% & 92.0 \\ \hline
        12 & 60.0 & 318.0  & 84.0 & 112.0 & 573.91 & 39.29 \%&  65.0\\ 
        
        \hline
        \hline
    \end{tabular}
    }
\vspace{-0.2cm}
\end{table}

In summary, the interaction counts and location heatmaps corroborate the sociograms by reflecting the behavior of participation and movement observed in speaking, proximity, and shared attention. High interaction counts align with dominant roles seen in the sociograms, while clustered or dispersed spatial patterns support the varying levels of group cohesion represented in the proximity and shared attention graphs.

\subsubsection{Task Performance}
\label{sec:task-performance}
This section evaluates task performance by analyzing key metrics such as distinct image labels, label changes, and group interactions to understand the impact of decision-making and collaboration.
Group-level task metrics in~\autoref{table:group-level-task-metrics} show participant interaction, task execution, and overall performance. The number of images grabbed and total image grabs, ranging from $50$ to $72$ and $220$ to $497$, respectively, indicate group activity levels. 
Higher counts, like in Group 3 ($71$ images, $497$ grabs), indicate greater involvement, while lower counts in Groups 1 and 9 suggest more restrained participation. 
Image label overrides, varying from $22$ to $44$, highlight decision revisions, with higher numbers (Group 7's $44$ overrides) indicating potential indecision or collaboration. Groups generally examined $110$--$112$ images, enabling fair comparisons. 
Distinct labels and label changes further show behavior diversity; Groups 3 and 11's highly distinct labels ($71$ and $72$) suggest diverse strategies, while Group 7's $101$ changes point to active debate.

Completion time and subjective accuracy are primary performance indicators. Outlier analysis found Group 5 with a low completion time ($209.27$ seconds) and Group 8 with a high time ($994.05$ seconds) and low accuracy ($17.86\%$), illustrating variability and a speed-accuracy trade-off. Participant 8D’s high interaction count, shown in~\autoref{fig:image-interactions}, supports this. In contrast, Group 1, with moderate interaction ($415.54$ seconds), achieved higher accuracy ($67.86\%$), indicating efficient performance. 
This aligns with~\cite{yangImmersiveCollaborativeSensemaking2022}, which reports average group accuracy below 60\%, highlighting the subjective nature of emotion-based sorting. A Pearson correlation analysis revealed a moderate negative relationship between completion time and accuracy ($r=-0.61$), hinting at a speed-precision trade-off.
We provide further details in Appendix~\ref{appendix:sociograms}, \autoref{fig:grouping-interactions} shows visual heatmaps illustrating group task performance, and~\autoref{fig:user-list-image} displays individual image interactions showing attention focus and participant engagement.

The analysis reveals that collaborative strategies and participant interactions affect task outcomes. Groups with balanced participation and cohesive decision-making (Group 11) performed better. Conversely, groups with fragmented behavior or extensive deliberation (Groups 7 and 8) often struggled and performed poorly.

\begin{table}[]
\caption{High-level group behavior label across all three types of sociogram.
}
\vspace{-0.35cm}
\label{tab:group-label}
\small
\begin{tabular}{|c|c|c|}
\hline
\textbf{Fragmented} & \textbf{Cohesive} & \textbf{Competitive}\\\hline
Groups 1, 2, 5, 6, 7, 8, 10, and 12 & Groups 4 and 11 & Groups 3 and 9\\ \hline
\end{tabular}
\end{table}
\subsection{Group Behavior Profiling Through Sociometric Analysis}
\label{sec:group-profiling}
This section analyzes graph attributes across three sociograms (conversation, proximity, and attention).

\subsubsection{Group Behavior Labels Generated in \sysname}
We categorize group behavior as \textbf{cohesive, competitive, and fragmented} labels based on aggregated characteristics across all sociograms, shown in~\autoref{tab:group-label}. 
Fragmented groups exhibit weaker internal connections, indicating lower alignment and coordination among members across sociograms. These groups show more dispersed interactions, suggesting variability in engagement and decision-making processes.
Cohesive groups (4 and 11) demonstrate strong internal connectivity and stability across sociograms, with members maintaining consistent engagement and alignment in interactions. These groups exhibit balanced participation, reinforcing mutual understanding and collaborative decision-making.
Competitive behavior, observed in Groups 3 and 9, indicates an interaction pattern in which members make assertive, possibly conflicting contributions.
This classification highlights how different sociogram weightings influence emergent group behaviors, offering insights into optimizing collaboration dynamics in MR environments.

\subsubsection{Graph Metrics Influence in Group Label Assignment}
\begin{figure}[t]
  \centering
  \begin{subfigure}[b]{0.49\linewidth}
    \centering
    \includegraphics[width=\linewidth]{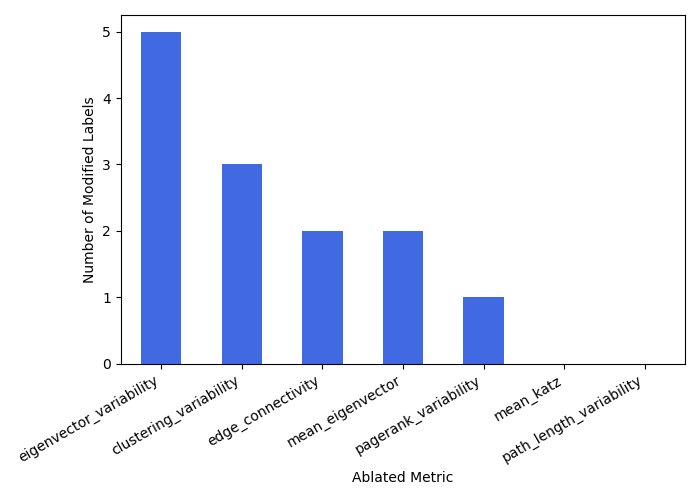}
    \vspace{-0.2cm}\caption{ }  
    \label{fig:ablation_bar}
  \end{subfigure}
  \vspace{-0.2cm}
  \hfill
  \begin{subfigure}[b]{0.49\linewidth}
    \centering
    \includegraphics[width=\linewidth]{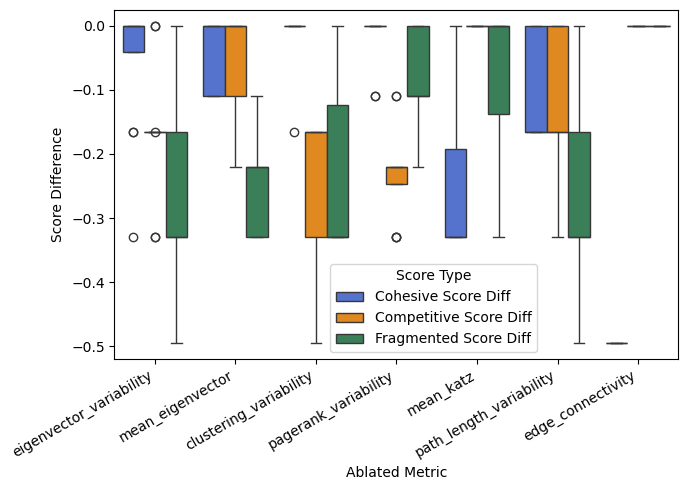}
    \vspace{-0.2cm}\caption{ }  
    \label{fig:ablation_box}
  \end{subfigure}
  \vspace{-0.2cm}
  \caption{Ablation Study Results: (a) Bar plot showing the number of labels affected when each graph metric is removed, and (b) Box plots depicting the distribution of group label scores for each metric removal scenario.}
  \vspace{-0.4cm}
  \label{fig:ablation}
\end{figure}
To understand the impact of graph metrics on \sysname's group labels, we conducted an ablation study as shown in~\autoref{fig:ablation}, analyzing each metric’s contribution to classification. Eigenvector variability had the strongest influence, altering labels for five of twelve groups (\autoref{fig:ablation_bar}), underscoring its role in capturing group dynamics. 
\autoref{fig:ablation_box} illustrates the score difference, showing how removing a metric shift computed label scores even if final classifications remain unchanged. While Mean Katz centrality and path length variability did not alter final labels, they significantly impacted label scores, indicating their role in refining classification boundaries. The varying effects of metric removal across cohesive, competitive, and fragmented groups provide insights into how each metric shapes behavioral categorization. This study clarifies \sysname’s internal mechanics and informs refinements for more precise group behavior assessment. These findings align with our statistical analysis of graph structure relationships with the final group label as shown in Appendix~\ref{sec:appx:key-graph-att:analysis}. A stability analysis of \sysname using Monte Carlo simulations with $±5\%$ random noise showed $98.17\%$ label consistency across 1000 iterations, confirming the algorithm's robustness to minor perturbations in sociogram edge weights. See Appendix~\ref{sec:appx:stability_analysis} for detailed results and refinements.

\begin{figure}
    \centering
    \includegraphics[width=0.6\linewidth]{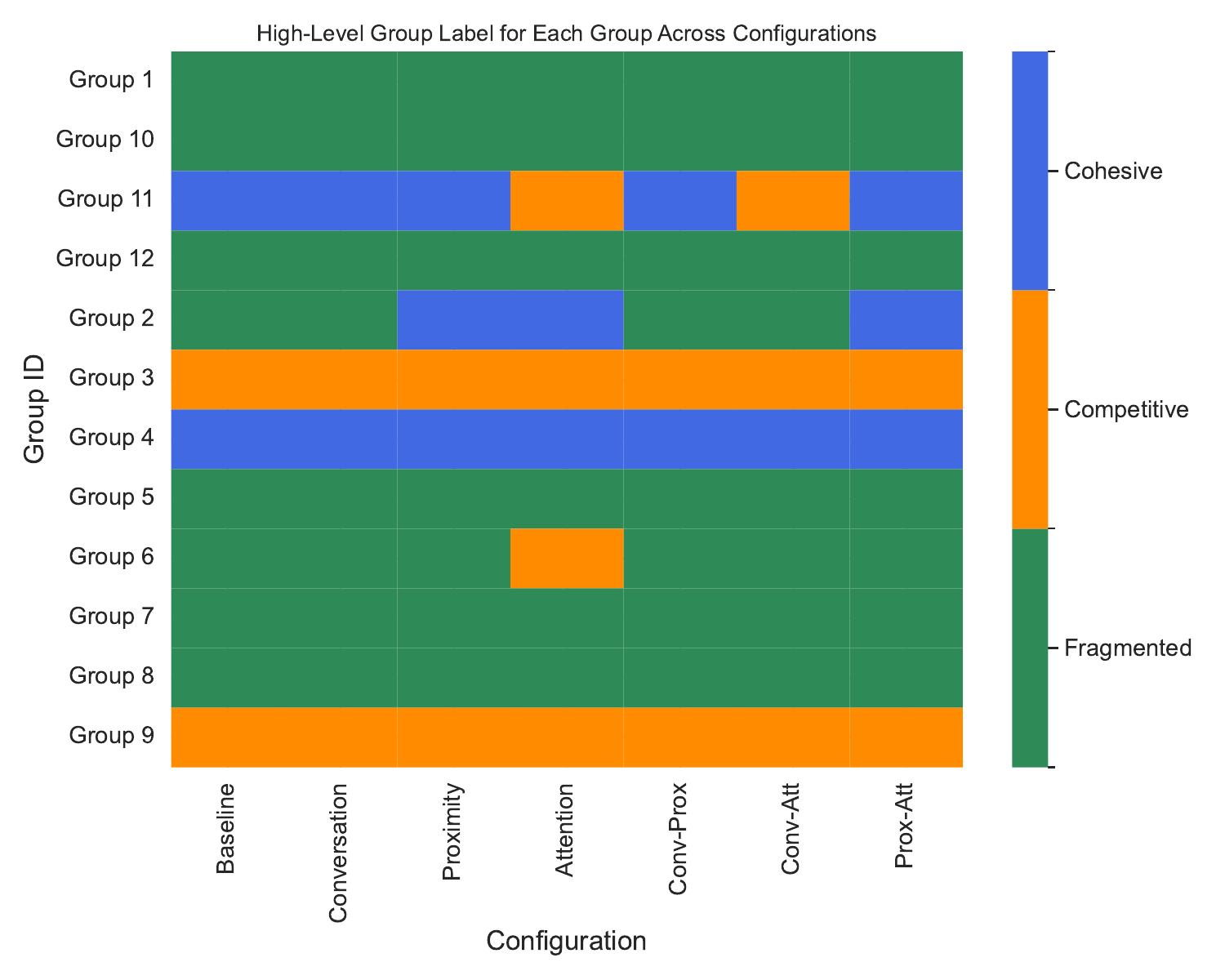}
    \vspace{-0.4cm}
    \caption{Heatmap displaying each group's high-level group label assignments across configurations. Each cell represents the assigned group behavior label for a specific group and configuration, visually mapping consistency or label variation. }
    \vspace{-0.3cm}
    \label{fig:group_label_heatmap}
\end{figure}
\subsubsection{Analysis of Group Behavior Distribution}
\label{sec:group-behvaior-distrbution}
We evaluated the impact of different sociogram weight combinations on group behavior predictions, testing configurations ranging from equal weighting to skewed distributions emphasizing specific interaction types. For example, a conversation-focused setup assigned 0.5 weight to conversation and 0.25 to others, while balanced setups like conversation-proximity (0.4 each, 0.2 for shared attention) explored dual-context effects. \autoref{fig:group_label_heatmap} visually maps the prevalence of each group's behavior across configurations, with cell intensity reflecting the strength of each group's behavior. The results confirm that configuration choices influence which group states emerge as dominant. Shared attention-focused setups led to more competitive groups, while proximity-emphasized configurations showed increased cohesiveness, though not dominantly. Balanced configurations generally behaved similarly to the baseline configurations. The results confirm that weighting choices shape dominant group states. See \autoref{fig:dist-group-level} and \autoref{fig:group-label-average-score} in Appendix~\ref{sec:appx:group-behvaior-distrbution} for label distributions and score variations, showing how interaction influences group dynamics and guiding model optimization.

\subsection{Statistical Analysis}
\label{sec:stat-eval}
This section presents the descriptive and statistical analyses of various metrics derived from the presence (PQ and IPQ) questionnaires, NASA TLX scores, and our custom group behavior survey. These metrics are analyzed to understand the differences between cohesive and fragmented groups and their impacts on subjective experience. 
\begin{table}[t]
\scriptsize
\caption{Mean scores per group for Presence (PQ+IPQ) Score, TLX Score, and group-related metrics, where group cohesion, attention, proximity, conversation, and collaboration represent the interpersonal dynamics and collaborative interactions.}
\label{tab:per-group-survey-scores}
\vspace{-0.3cm}
\begin{tabular}{|c|c|c|ccccc|}
\hline
\hline
\multirow{2}{*}{\textbf{Group ID}} & \multirow{2}{*}{\textbf{Presence Score}} & \multirow{2}{*}{\textbf{TLX Score}} & \multicolumn{5}{c|}{\textbf{Group Behavior  Metrics}} \\ \cline{4-8} 
 &  &  & \multicolumn{1}{c|}{\textbf{Cohesion}} & \multicolumn{1}{c|}{\textbf{Attention}} & \multicolumn{1}{c|}{\textbf{Proximity}} & \multicolumn{1}{c|}{\textbf{Conversation}} & \textbf{Collaboration} \\ \hline
 \hline
\textit{\textbf{1}} & 4.86 & 2.2 & \multicolumn{1}{c|}{5.75} & \multicolumn{1}{c|}{4.25} & \multicolumn{1}{c|}{5.5} & \multicolumn{1}{c|}{6.25} & 6.25 \\ \hline
\textit{\textbf{2}} & 5.04 & 2.55 & \multicolumn{1}{c|}{6.75} & \multicolumn{1}{c|}{5.25} & \multicolumn{1}{c|}{4.75} & \multicolumn{1}{c|}{6.25} & 6.75 \\ \hline
\textit{\textbf{3}} & 4.5 & 1.9 & \multicolumn{1}{c|}{7} & \multicolumn{1}{c|}{4} & \multicolumn{1}{c|}{2.25} & \multicolumn{1}{c|}{7} & 6.75 \\ \hline
\textit{\textbf{4}} & 5.32 & 2.55 & \multicolumn{1}{c|}{6} & \multicolumn{1}{c|}{5} & \multicolumn{1}{c|}{5} & \multicolumn{1}{c|}{6.25} & 6.25 \\ \hline
\textit{\textbf{5}} & 5.22 & 1.2 & \multicolumn{1}{c|}{7} & \multicolumn{1}{c|}{4.75} & \multicolumn{1}{c|}{4} & \multicolumn{1}{c|}{5.25} & 7 \\ \hline
\textit{\textbf{6}} & 5.4 & 1.95 & \multicolumn{1}{c|}{7} & \multicolumn{1}{c|}{5} & \multicolumn{1}{c|}{3.5} & \multicolumn{1}{c|}{6.5} & 7 \\ \hline
\textit{\textbf{7}} & 5.07 & 2.3 & \multicolumn{1}{c|}{7} & \multicolumn{1}{c|}{7} & \multicolumn{1}{c|}{3.5} & \multicolumn{1}{c|}{6.75} & 7 \\ \hline
\textit{\textbf{8}} & 5.06 & 2.5 & \multicolumn{1}{c|}{6.5} & \multicolumn{1}{c|}{5.75} & \multicolumn{1}{c|}{3} & \multicolumn{1}{c|}{6.5} & 6.75 \\ \hline
\textit{\textbf{9}} & 4.75 & 2.25 & \multicolumn{1}{c|}{6.5} & \multicolumn{1}{c|}{5.5} & \multicolumn{1}{c|}{3.75} & \multicolumn{1}{c|}{6.25} & 5.75 \\ \hline
\textit{\textbf{10}} & 4.81 & 2.95 & \multicolumn{1}{c|}{6.25} & \multicolumn{1}{c|}{5} & \multicolumn{1}{c|}{4} & \multicolumn{1}{c|}{5.5} & 6.75 \\ \hline
\textit{\textbf{11}} & 4.15 & 2.9 & \multicolumn{1}{c|}{6.25} & \multicolumn{1}{c|}{5.25} & \multicolumn{1}{c|}{4} & \multicolumn{1}{c|}{5.75} & 6.75 \\ \hline 
\textit{\textbf{12}} &4.50 &  2.70& \multicolumn{1}{c|}{6.00 } & \multicolumn{1}{c|}{6.00 } & \multicolumn{1}{c|}{3.75} & \multicolumn{1}{c|}{5.50} & 6.00 \\ \hline \hline
\end{tabular}
\vspace{-0.3cm}
\end{table}

~\autoref{tab:per-group-survey-scores} shows per-group means for presence, TLX, and group behavior with a summary of descriptive statistics, showing participants' subjective experiences and response variability. Metrics like PQ ($\mu 5.46$) and IPQ ($\mu 4.39$) indicate moderate presence levels. Groups with higher presence scores, like Groups 2 and 6, reported strong cohesion and collaboration, linking presence to effective teamwork. Fragmented groups, like Group 11, had lower proximity and attention scores, reflecting challenges in maintaining engagement and shared focus. Notably, other groups showed even lower proximity and attention scores than Group 11, yet their overall group cohesion remained stable. Group 3, for example, had low proximity (2.25) but maintained a high cohesion score (7.0), suggesting that proximity alone is not always a predictor of fragmentation and that other factors, such as conversation and collaboration scores, contribute to maintaining cohesion. Similarly, Group 8, despite a low proximity score (3.0), exhibited strong attention (5.75) and conversation (6.5), reinforcing that cohesion can emerge from different interaction patterns, not just physical closeness. The results highlight the complex relationship between group behavior metrics and subjective experiences. While cohesive groups tend to report higher presence and collaboration, fragmented groups do not always follow a linear pattern of disengagement. Instead, variations in proximity, attention, and conversation suggest that different interaction styles contribute uniquely to perceived group effectiveness.
For a detailed summary of presence and IPQ subscales, refer to \autoref{tab:survey-summary-stats} in Appendix~\ref{sec:appx:behavioral-measure}. These findings emphasize the need for a nuanced approach to interpreting group behavior, where multiple metrics collectively define the dynamics of cohesion, fragmentation, and competitiveness in collaborative environments.

\subsection{Mediation Analysis}
\label{sec:mediation-analysis}
\begin{figure}
    \centering
    \vspace{-0.3cm}
    \includegraphics[width=.7\linewidth]{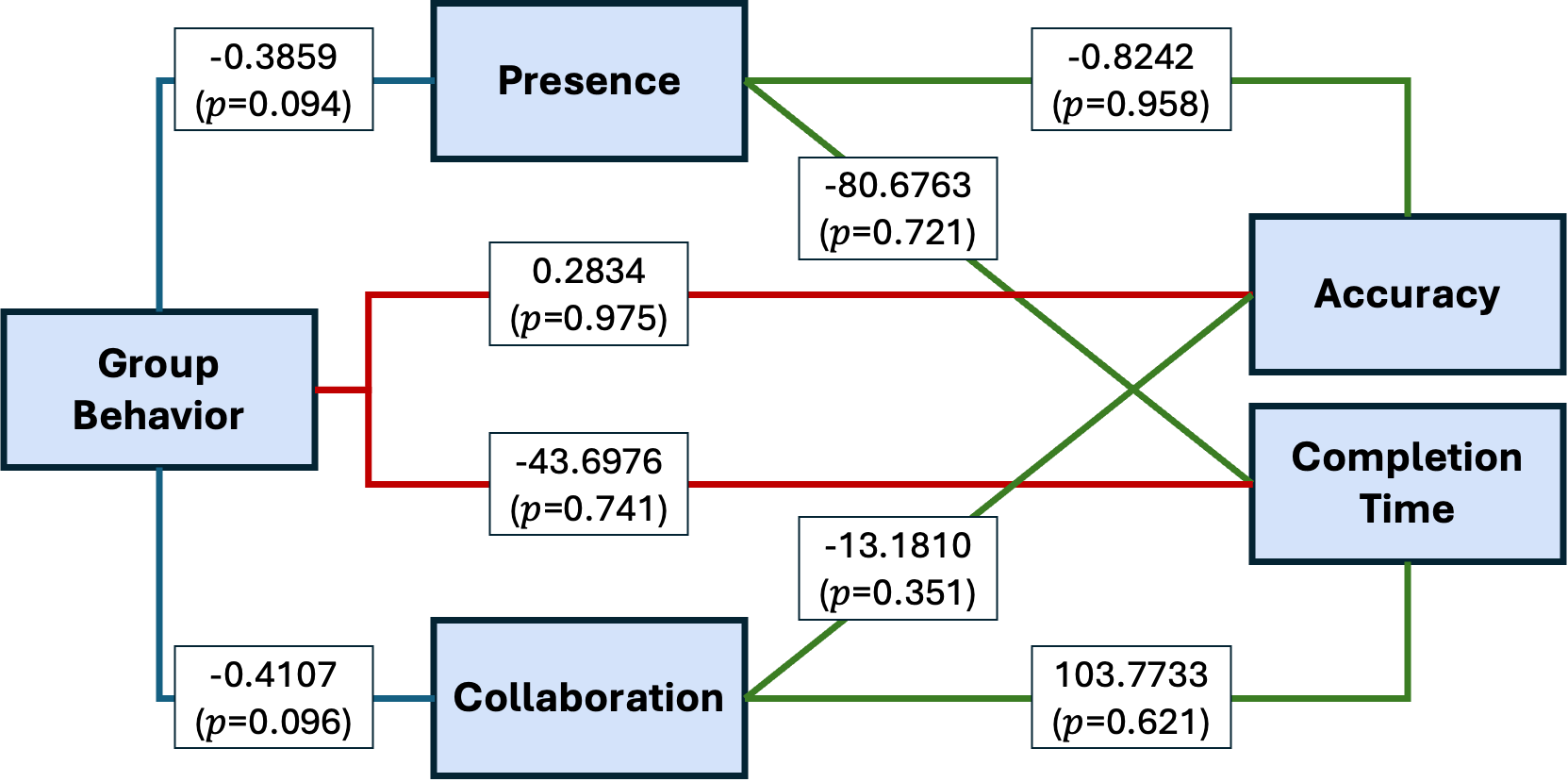}
    \vspace{-0.3cm}
    \caption{Mediation analysis between group behavior and task performance, mediated by individual behavior. 
    }
    \vspace{-0.1cm}
    \label{fig:mediation}
\end{figure}
We conducted a mediation analysis to examine whether group behavior influences task performance, measured by completion time and accuracy, through presence and collaboration as mediators. The goal was to determine whether group behavior indirectly affects task outcomes via individual experience (presence) and group interaction (collaboration), both captured through sociograms.
Presence and collaboration were chosen as mediators due to their higher correlations with the variables. Ordinary least squares regression models were used to quantify direct and indirect effects. Appendix~\ref{sec:task-behavior-metrics-vs-group} shows that task performance differences across group types are minimal, prompting our mediation analysis to examine indirect effects through presence and collaboration. The results show no significant direct or indirect effects of group behavior on task performance. The direct path from group behavior to accuracy was non-significant, and neither presence nor collaboration significantly mediated accuracy outcomes. The path from group behavior to presence showed a moderate negative effect ($-0.38$, $p=0.094$) with borderline significance, but its influence on completion time was non-significant ($-80.67$, $p=0.72$), leading to an overall non-significant indirect effect ($31.14$) and a proportion mediated of $-0.71$. A similar trend appeared for collaboration, where its effect on completion time was positive but non-significant ($103.77$, $p=0.621$), yielding an indirect effect of $-42.62$ and a proportion mediated of $0.97$. For accuracy, neither presence nor collaboration showed significant mediation effects. The direct path from group behavior to accuracy was also non-significant. These results indicate that while presence and collaboration might mediate the effect on completion time, their roles are not statistically significant and need further study to clarify their impact on group dynamics.

\emph{This result summarizes the study: task performance does not directly shape group behavior, either through individual experience or interpersonal dynamics. Instead, group behavior emerges independently of measured performance outcomes, suggesting that social interaction patterns are not necessarily driven by task efficiency or accuracy. While strong collaboration and presence may enhance subjective experience, they do not always align with improved task performance, indicating that external factors such as individual cognitive load, prior experience, or external constraints may play a more significant role. 
Although this result does not establish a direct causal link, it provides a key insight. \emph{Group behavior is influenced by complex social and cognitive factors rather than just task performance metrics}, underscoring the need for further research to explore how external and individual factors interact with group behavior beyond performance-driven metrics.}

\section{Discussion}\label{sec:discuss}

\subsection{Characterizing Group Behavior}
\label{sec:intrepreting-results}
\sysname’s group labels \texttt{cohesive, fragmented, and competitive} capture collaboration dynamics across conversation, proximity, and shared attention, aligning with trends in task performance, subjective experience, and statistical outcomes. Cohesive groups, such as Groups 4 and 11, exhibited balanced interactions, strong shared attention, and stable proximity across sociograms, reinforcing high collaboration and presence scores. Their conversation sociograms showed distributed participation, with no dominant speakers, aligning with their high collaboration and cohesion scores in the survey results. Proximity and shared attention sociograms confirmed their strong engagement, with members maintaining spatial closeness and shared focus, further supporting their classification. Statistical analysis reinforced this pattern, with cohesive groups reporting higher presence and collaboration scores, indicating stronger group alignment.
Fragmented groups displayed uneven participation, dominant speakers, and weaker proximity. Their conversation sociograms showed centralized or highly skewed speaking patterns, often with one or two members driving discussions while others contributed minimally. Proximity sociograms reflected looser physical clustering, while shared attention graphs showed less sustained focus, contributing to lower engagement. These patterns align with their lower presence and collaboration scores, reinforcing their classification as fragmented. Statistical tests further support this distinction, with fragmented groups showing more variation in attention and proximity scores, indicating inconsistent engagement.
Competitive groups showed centralized influence in their sociograms but varied levels of conversation balance and proximity. While their conversation graphs displayed structured interactions, influence was concentrated among a few members, suggesting a hierarchical dynamic. Proximity sociograms revealed mixed spatial engagement, indicating that while these groups maintained structured coordination, their interactions were less evenly distributed than cohesive groups. The statistical analysis supports this classification, as these groups had moderate presence and collaboration scores, reflecting structured yet less distributed participation. 

While task metrics such as completion time and accuracy varied across groups, the more fundamental insights lie in the relationship between subjective experience and group behavior. The moderate negative correlation between completion time and accuracy suggests a speed-accuracy trade-off, where extended deliberation does not always yield better results. However, completion time and accuracy alone do not fully capture interaction quality. Balanced participation correlated with higher cohesion and collaboration, while dominant conversational contributions pointed to hierarchical structures that may limit collaboration. Proximity and shared attention graphs supported these findings, showing that strong physical closeness and shared attention indicate cohesive interactions.
The mediation analysis confirms that group behavior does not emerge as a direct consequence of task performance. Presence and collaboration varied across groups but did not significantly mediate performance outcomes such as completion time and accuracy, reinforcing that social dynamics arise from behavioral and cognitive interactions rather than task efficiency alone. Stability analysis further supports this, as most groups maintained consistent labels, but Group 3’s classification fluctuated, indicating its competitive behavior was near a decision boundary, sensitive to subtle interaction shifts.

Taken together, sociograms, statistical findings, mediation results, and stability analysis reinforce that cohesion, fragmentation, and competition emerge from distinct interaction structures rather than task efficiency alone. \sysname’s classification effectively captures these behavioral distinctions, demonstrating how different interaction patterns shape collaboration beyond performance-driven measures.

\subsection{Implications}
\label{sec:impication}
The results demonstrate that group behavior is shaped by cognitive, spatial, and social factors rather than direct performance outcomes, emphasizing the need for systems that capture these interactions dynamically. \sysname provides a structured approach to assessing group cohesion, fragmentation, and competitiveness, offering a scalable method for analyzing real-time collaboration patterns in MR environments. By continuously measuring interaction metrics, \sysname enables a deeper understanding of how social and behavioral dynamics evolve during collaborative tasks, supporting the design of interventions that will promote more effective teamwork.

The distinction between \texttt{cohesive, fragmented, and competitive} groups suggests that collaboration strategies should not focus solely on optimizing task efficiency but also on promoting engagement, shared focus, and equitable participation. The strong alignment between sociometric indicators and subjective experience underscores the importance of designing collaborative systems that enhance presence and coordination rather than just improving performance outcomes. Fragmented groups exhibited signs of disengagement and uneven participation, pointing to the need for adaptive support mechanisms that mitigate imbalances in interaction. Strategies such as real-time feedback on group engagement, role rebalancing mechanisms, or adaptive spatial prompts could help maintain cohesion and prevent exclusionary dynamics.

The findings also highlight the broader impact of social structures on collaborative experiences, reinforcing the role of presence and engagement in shaping group interactions. While hierarchical organization in competitive groups can enhance efficiency, it may limit distributed decision-making, suggesting that system design should allow flexibility in structuring group roles. Spatial and attentional alignment emerged as key factors in maintaining cohesive interactions, indicating that MR environments should be designed to encourage shared attention and equitable conversational dynamics.
Beyond MR, these insights extend to the design of remote and hybrid collaboration platforms, where real-time group behavior monitoring could improve interaction quality. The statistical results suggest that presence and collaboration, though not direct mediators of task success, significantly influence perceived engagement, making them essential considerations for optimizing collaborative environments. Future research should refine sociometric models to better capture subtle behavioral shifts and explore interventions that promote balanced group engagement across diverse contexts. By integrating behavioral insights with real-time system adaptations, \sysname provides a foundation for advancing collaborative intelligence in MR and beyond.

\section{Limitation and Future Work}\label{sec:limitations}
\sysname provides a structured approach to understanding group behavior, but some limitations must be considered. First, while shared attention is a key metric, it may not fully capture collaboration in neurodivergent individuals, such as those on the autism spectrum, who exhibit different joint attention patterns~\cite{caruanaJointAttentionDifficulties2018}. Future work should explore broader inclusivity in behavioral modeling. Second, \sysname currently relies on proximity, conversation, and shared attention, but expanding sensor modalities such as facial expressions and physiological responses could provide richer insights into group interactions. Third, while this study focuses on fixed group behavior within a single task, real-world collaboration involves evolving group dynamics. Integrating dynamic network analysis and social psychology theories, such as Tuckman’s group development model~\cite{tuckman1965developmental}, could enhance understanding of interaction shifts over time. Additionally, the findings are based on a limited sample size and a single collaborative task, restricting generalizability. Future studies should include larger, more diverse groups and a broader range of tasks to strengthen \sysname~’s applicability across various collaborative settings. Finally, while \sysname quantifies interaction patterns, it does not capture nuanced subjective experiences such as frustration, motivation, or perceived collaboration quality. Incorporating qualitative feedback through participant reflections or interviews would provide deeper insights into how individuals experience group interactions. A more detailed discussion of these limitations and future research directions is provided in Appendix~\ref{sec:appx-limitations}.

\section{Conclusion}\label{sec:conclusion}
This study advances our understanding of group behavior in MR by introducing \sysname, a novel framework integrating multiple data sources and analysis techniques.  \sysname uniquely combines these elements to analyze group behavior within the MR context, capturing previously difficult-to-measure nuanced interaction patterns. This provides new insights into how MR environments influence group behavior and performance, surpassing traditional methods and purely virtual settings.  This work aims to facilitate the development of more effective collaborative MR systems.
Specifically, \sysname analyzes collaborative MR tasks through passive sensor data, modeling pairwise interactions in conversation, shared attention, and proximity.  Sociometric analysis reveals group behaviors like cohesion, competitiveness, and fragmentation. A user study demonstrated its effectiveness in identifying dominant group behaviors. While individual differences and task traits influence behaviors, \sysname offers a valuable tool for analyzing collaborative MR experiences. Our evaluation shows that group behavior in MR emerges independently of task performance, emphasizing the role of interaction dynamics over efficiency metrics. As XR technologies advance, such frameworks will enhance collaboration and optimize performance in these emerging realities.

\begin{acks}
This work is supported by the U.S. National Science Foundation (NSF) under grant number 2339266.

\end{acks}

\bibliographystyle{ACM-Reference-Format}
\bibliography{paper}

\clearpage
\appendix
\vspace{0.2cm}
\noindent\begin{LARGE}\textbf{Appendix} \vspace{4mm} \end{LARGE}

We provide additional information for our paper, \emph{\sysname: Interpreting Group Behavior in Mixed Reality}, in the following order:

\begin{itemize}
    \item Additional Related Work (Appendix~\ref{sec:appx:related})
    \item Data Processing Module Implementation (Appendix~\ref{sec:appx:data-proc})
    \item Theoretical Justification of the Group Behavior Characterization Module (Appendix~\ref{sec:appx:characterization})
    \item Terminology (Appendix~\ref{sec:appx:terms})
    \item Pilot Study (Appendix~\ref{app:pilot})
    \item Additional Experimental Results (Appendix~\ref{sec:appx:additional-results})
    \item Additional Limitation Details (Appendix~\ref{sec:appx-limitations})
\end{itemize}

\section{Additional Related Work}
\label{sec:appx:related}

\subsection{Sensing Individual and Group Behavior in Ubiquitous Systems}\label{sec:appx:sensing_behavior}
Recent advances in ubiquitous sensing systems have enabled the precise capture of individual and group behaviors across diverse environments~\cite{jarin2023behavr}. Physiological sensing, in particular, has been extensively applied to understand affective states, especially in high-stress contexts. Wearable devices such as heart rate monitors, electrodermal activity (EDA) bands, and EEG headbands capture emotional workload and stress responses, offering insights into real-time emotional regulation~\cite{parkHideseekDetectingWorkers2024, laporteExploringLAUREATELongitudinal2023, gebhardtDetectingUsersEmotional2024} and adaptive education environment~\cite{taherisadr2023erudite}. In social environments, these sensors track interpersonal synchrony, capturing shared engagement between individuals and groups, such as actors and audiences in live performances~\cite{sunUsingWearableSensors2023}.

Additionally, these technologies have been deployed to monitor social interactions and engagement in educational and collaborative settings. Systems like CoCo utilize audio-video sensors to analyze non-verbal communication cues during video conferencing, thereby assessing team dynamics~\cite{samroseCoCoCollaborationCoach2018}. Wearable sensors also facilitate the capture of group cohesion in small teams by measuring physical movement, proximity, and mimicry patterns~\cite{zhang_teamsense_2018}. Such sensing technologies enhance our ability to observe and interpret human behavior in real-time, supporting adaptive interventions across collaborative, educational, and affective computing contexts.

In VR, sensors embedded in headsets and controllers track user movements, gestures, and spatial positioning to analyze collaboration dynamics~\cite{yangImmersiveCollaborativeSensemaking2022, letarnecImprovingCollaborativeLearning2023}. Additionally, prior work integrated sensors into VR head-mounted displays (HMDs) to analyze physiological signals, providing insights into individual states~\cite{chiossiSensConEmbeddingPhysiological2023}.  Similarly, volumetric capture through depth cameras enables the creation of 3D avatars, allowing real-time monitoring of interactions, proxemics, and non-verbal cues in MR settings~\cite{irlittiVolumetricMixedReality2023}. To our knowledge, no research has yet explored passive and automated group behavior characterization in MR with sensors in modern headsets and this work aims to fill that gap.

\begin{figure}
  \centering
  \includegraphics[height=8.4cm, trim={0 4.5cm 0 0},clip]{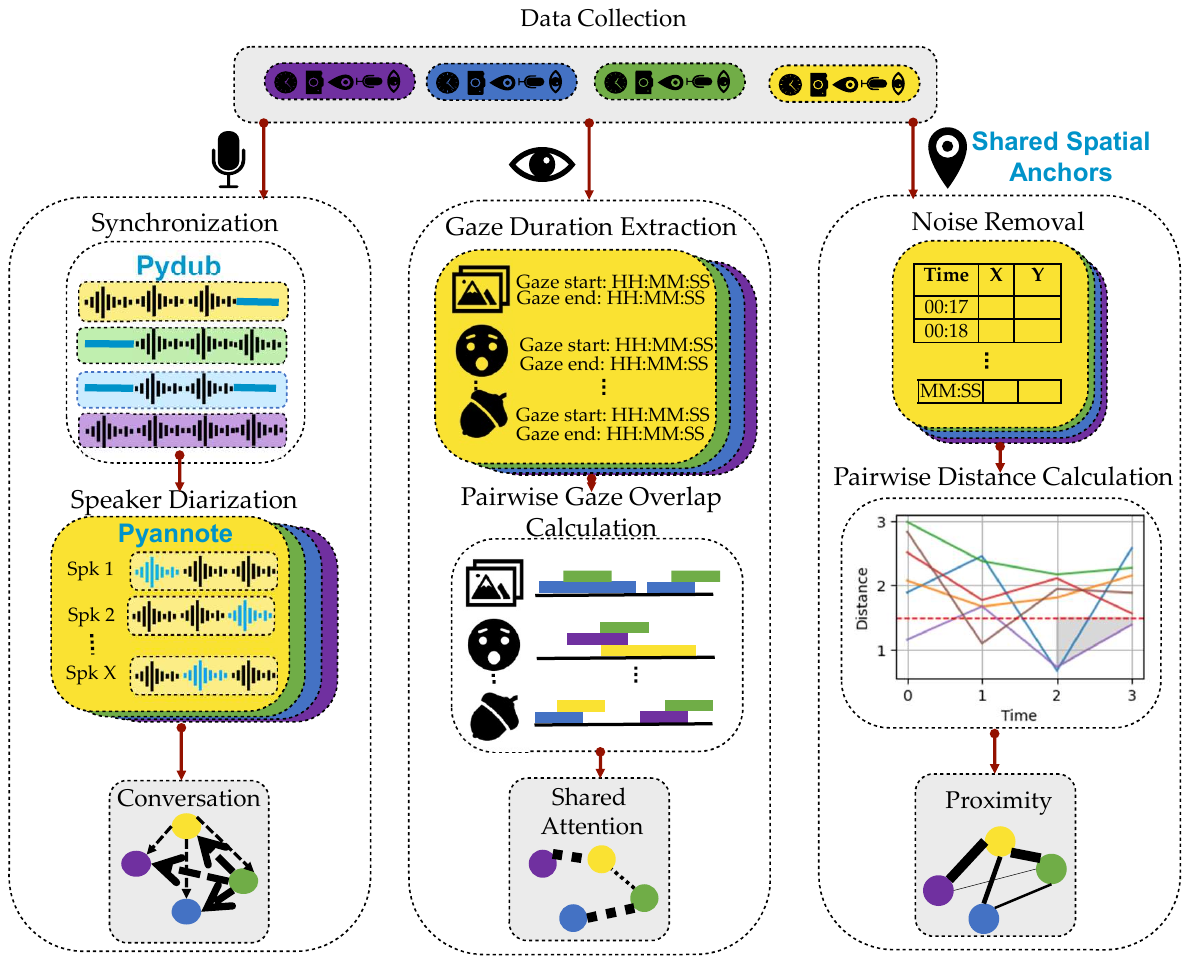}
  \vspace{-0.2cm}
  \caption{Overview of \sysname processing module to model raw sensor data into sociograms, depicting various interaction types, such as conversation, shared attention, and proximity.}
  \vspace{-0.3cm}
  \label{fig:data-processing-summary}
\end{figure}

\section{Data Processing Module Implementation}\label{sec:appx:data-proc}
A summary of the data processing pipeline for each of the sociograms can be seen in \autoref{fig:data-processing-summary}.

\subsection{Conversation Sociogram Data Processing Pipeline}\label{sec:appx:conv=data-proc}
The data processing pipeline for generating conversation sociograms involves several steps. We began by collecting raw audio data using the built-in microphones in the MR headsets, with timestamps recorded at a sampling rate of $44,100$ \texttt{Hz}. We synchronized the audio recordings and padded shorter ones using the Pydub library~\cite{pydub} to ensure consistent timestamps across all participants. For segmenting conversations, we utilized the speaker diarization model, Pyannote~\cite{pyannote} to segment recordings into individual speech turns. We identified speakers by extracting speaker embeddings and clustering them based on acoustic features.  The speaker with the loudest average volume was identified as the primary speaker. The speaker with the loudest average volume was identified as the primary speaker, assuming they were closest to the microphone. This process results in a conversation sociogram: a directed graph where nodes represent participants, edges denote conversation direction, and edge weights reflect interaction duration in seconds.

\subsection{Shared Attention Sociogram Data Processing Pipeline}\label{sec:appx:gaze-proc}
To analyze raw eye-tracking data and generate the shared attention sociogram, we used built-in sensors in the MR headset, collecting data with Meta's Eye Tracking SDK for Unity ~\cite{noauthor_eye_nodate}. A custom script logged the start and end times of participants' eye gaze intersecting with virtual objects. We parsed these logs to isolate gaze durations for each user on virtual objects, timestamping each gaze event to determine precise interaction durations. This allowed for a detailed analysis of gaze patterns and interactions within the virtual environment. We then implemented overlap calculation to identify instances of shared focus on the same virtual object within a shared time frame. Overlap durations exceeding 13 milliseconds were aggregated. Using these gaze durations, we generated an undirected gaze-based interaction sociogram. In this sociogram, nodes represent individual participants, edges represent shared attention on virtual objects, and edge weights indicate the cumulative duration of shared gaze on mutual virtual objects. This approach ensures that the sociogram accurately reflects significant interactions and shared focus in the MR environment.

\subsection{Proximity Sociogram Data Processing Pipeline} \label{sec:appx:prox-proc}
To analyze spatial data and generate the proximity sociogram, we implemented a data processing pipeline~\autoref{fig:data-processing-summary}. We recorded each participant's headset location every second and synchronized the coordinate system across all participants using Shared Spatial Anchors via Photon Unity Networking ~\cite{noauthor_shared_nodate}. This ensured spatial data consistency. We synchronized timestamps and excluded the first and last 15 seconds of data to remove noise from the application loading and exiting. For each timestamp, we calculated Euclidean distances between all participant pairs.

\begin{table}[h]
\centering
\small
\renewcommand{\arraystretch}{1.2}
\caption{Mapping of Graph Metrics to Group Scores and Labels}
\begin{tabular}{|c|c|c|p{2cm}|}
\hline
\textbf{Metric} & \textbf{Score Category} & \textbf{Characteristic Value} & \textbf{Contribution to Label} \\ \hline

\textbf{Eigenvector Centrality Variability} ($\sigma_{x_i}$) & \multirow{2}{*}{Cohesion} & High, Distributed, Tight-knit & \textbf{Cohesive} \\ \cline{1-1} \cline{3-4}
\textbf{Clustering Coefficient Variability} ($\sigma_{C_i}$) &  & Low, Fragile, Loose-knit & \textbf{Fragmented} \\ \hline

\textbf{Eigenvector Centrality Mean} ($\mu_{x_i}$) & \multirow{3}{*}{Influence} & High & \textbf{Cohesive} \\ \cline{1-1} \cline{3-4}
\textbf{PageRank Variability} ($\sigma_P$) &  & Moderate & \textbf{Competitive} \\ \cline{1-1} \cline{3-4}
\textbf{Katz Centrality Mean} ($\mu_{katz}$) &  & Low & \textbf{Fragmented} \\ \hline

\textbf{Edge Connectivity} ($\kappa'(G)$) & \multirow{2}{*}{Connectivity} & Resilient, Moderate, Centralized & \textbf{Competitive} \\ \cline{1-1} \cline{3-4}
\textbf{Path Length Variability} ($\sigma_{PL}$) &  & Low, Fragile, Loose-knit & \textbf{Fragmented} \\ \hline

\textbf{Weighted Score Contributions} & \multicolumn{2}{c|}{Final Group Label} & \textbf{Cohesive, Competitive, Fragmented} \\ \hline
\end{tabular}
\label{tab:group_behavior_metrics}
\end{table}

\section{Theoretical Justification of the Group Behavior Characterization Module}
\label{sec:appx:characterization}
\subsection{Group Behavior Characterization}
\label{sec:appx:group-behavior-characterization}
Our methodologies and the mathematical formulation of our selected metrics are summarized in~\autoref{alg:metric-computation}. The function $ComputeMetrics(G)$ in~\autoref{alg:metric-computation} computes all the metrics for a given sociogram graph $G$. The $G$ is represented by an adjacency matrix $A_{ij}$, where $A_{ij}$ contains the weight of the edge (interaction between two users) between nodes (user) $i$ and $j$ if no edge exists, $A_{ij} = 0$. $ComputeMetrics(G)$ function calculates the weighted graph metrics to evaluate nodes' and edges' structural significance for group behavior as summarized in ~\autoref{tab:group_behavior_metrics}.  These metrics quantify different aspects of interaction structures as follows:
\begin{itemize}
    \item Eigenvector Centrality (\$x\_i\$) and Clustering Coefficient (\$C\_i\$) assess \textbf{cohesion}, where higher values indicate strong, well-distributed group interactions, while lower values suggest fragmentation.
    \item PageRank (\$P(i)\$), Katz Centrality ($\mu\_{katz}$), and Eigenvector Centrality Mean ($\mu_{x_i}$) capture \textbf{influence}, identifying key participants who drive attention and interaction.
    \item Edge Connectivity ($\kappa'(G)$) and Path Length Variability ($\sigma\_{PL}$) measure \textbf{connectivity}, determining the resilience of the group and its ability to sustain interactions under structural changes.
\end{itemize}
After computing these metrics, as the first step, $GeneratePerInteractionCharacteristics$ in ~\autoref{alg:metric-computation}  processes metrics from each sociogram, pairs them according to the nature of the interaction, and aggregates them to identify high-level patterns of the group such as cohesion (a graph theory terminology), influence, and connectivity. These labels,\emph{ cohesion, influence, connectivity}, were chosen because they capture emergent properties of group behavior that cannot be understood by examining individual interactions in isolation. Group behavior arises from the collective patterns of interaction, shaping dynamics such as leadership, stability, and resilience. The function $GeneratePerInteractionCharacteristics(metrics)$ in~\autoref{alg:metric-computation} maps them to categorical scores:
\begin{itemize}
    \item Cohesion is classified as high, moderate, or low based on $\sigma_{C_i}$ and $\sigma_{x_i}$.
    \item Influence is assigned a score of high, moderate, or low using $\mu_{x_i}$, $\sigma_{P}$, and $\mu_{katz}$.
    \item Connectivity is categorized as resilient, moderate, or fragile, based on $\kappa'(G)$ and $\sigma_{PL}$.
\end{itemize}

Finally, these interaction-level scores contribute to the final group label (cohesive, competitive, or fragmented), determined through a weighted aggregation process using~\autoref{alg:final-group-label}. The algorithm systematically evaluates the structural role of members and their interactions, providing an interpretable classification of group behavior.

\setlength{\textfloatsep}{6pt}
\SetAlgoLined 
\begin{flushleft} 
\begin{algorithm}[t]
\small
\caption{\texttt{GroupCharacterizer}: Group Characterization using Graph Metrics}
\label{alg:metric-computation}
\SetKwInOut{KwIn}{Input}
\SetKwInOut{KwOut}{Output}
\SetKwFunction{LoadGraphs}{LoadGraphsFromPairwiseEdges}
\SetKwFunction{ComputeMetrics}{ComputeMetrics}
\SetKwFunction{GeneratePerInteractionCharacteristics}{GeneratePerInteractionCharacteristics}
\KwIn{Sociogram Graph $G = (V, E, W)$ for conversation, proximity, and shared attention data}
\KwOut{$group\_characteristics: cohesion, influence, connectivity, centralization, clustering$}

\SetKwProg{Fn}{Function}{:}{}
\Fn{\ComputeMetrics{$G$}}{
    Initialize $A_{ij} \gets$ Adjacency matrix of $G$, where $A_{ij} = \text{weight of edge } (i, j)$ if $(i, j) \in E$, otherwise $0$
\[
\text{parameters} \gets \{
\begin{array}{l l}
    \sigma_{st}: \text{Total number of shortest paths between } s \text{ and } t,\ \sigma_{st}(v): \text{Paths passing through } v,\\
     k_i: \text{Degree of } i ,\ \alpha: \text{Decay factor for Katz centrality } (\alpha < \frac{1}{\lambda_{max}}), \quad \beta: \text{Bias term for Katz},\\
     l_i: \text{Path length for node } i,\ 
    d: \text{PageRank damping factor } (d = 0.85), \quad N: \text{Total nodes in } G,\\
    M(i): \text{Nodes linking to } i,\ L(j): \text{Out-degree of } j,\ 
     e_i: \text{Edges between neighbors of } i,
    \\ \lambda: \text{Largest eigenvalue of }, S \gets \text{minimum set of E such that removal of S disconnects G}\;
\end{array}
\}
\]
    \If{$G$ is connected}{
       $eigenvector\_centrality \gets x_i = \frac{1}{\lambda} \sum_{j} A_{ij} x_j$\;
        $mean\_eigenvector\_centrality \gets \mu\_x_i  = \frac{1}{|V|} \sum_{i \in V} x_i$\;
        $eigenvector\_variability \gets \sigma\_x_i  =  \sqrt{\frac{1}{|V|} \sum_{i \in V} (x_i - mean\_eigenvector)^2}$\;
        $clustering\_coeff \gets C_i = \frac{2e_i}{k_i(k_i - 1)}$\;
        $clustering\_variability \gets \sigma\_C_i  = \sqrt{\frac{1}{|V|} \sum_{i \in V} (C_i - mean(C))^2}$\;
        $pagerank \gets P(i) = \frac{1-d}{N} + d \sum_{j \in M(i)} \frac{P(j)}{L(j)}$\;
        $pagerank\_variability \gets \sigma\_P = \sqrt{\frac{1}{|V|} \sum_{i \in V} (P(i) - mean(P))^2}$\;
        $mean\_katz \gets \mu_{katz} \text{ where } x_i = \alpha \sum_{j=1}^N A_{ij} x_j + \beta$\;
        $path\_length\_variability \gets \sigma\_{PL}\sqrt{\frac{1}{|V|} \sum_{i \in V} (l_i - mean(l))^2}$\;
        $edge\_connectivity \gets \kappa'(G) = \min_{S \subseteq E} |S|$
    }
\Return $metrics$ dictionary with computed values\;
}
\Fn{\GeneratePerInteractionCharacteristics{$metrics$}}{
    Initialize $scores \gets \text{default dictionary initialized to 0}$\;
     Initialize $group\_characteristics \gets \{cohesion, influence, connectivity\}$\;
   
    \ForEach{$(interaction\_type)$ in \{conversation, proximity, shared\_attention\}}{
$scores[cohesion] \gets \{high, moderate, low\}$ based on $\sigma\_C_i$ and $\sigma\_x_i$\;
$scores[influence] \gets \{high, moderate, low\}$ based on $\mu\_x_i$, $\sigma\_P$, and $\mu_{katz}$\;
$scores[connectivity] \gets \{resilient, moderate, fragile\}$ based on $\kappa'(G)$ and $\sigma\_{PL}$\;
}
    }
     \Return $group\_characteristics$\;
\end{algorithm}

\end{flushleft}

\subsection{Characterizing Cohesion, Influence, and Connectivity}
\label{sec:appx:charaterization}
Group dynamics in collaborative settings are shaped by how members interact, distribute influence, and maintain structural stability. To systematically analyze these aspects, we characterize group behavior through cohesion, influence, and connectivity, as these dimensions capture fundamental structural and functional properties of group interactions~\cite{watts1998collective-dynamics-cohesion}.

Cohesion reflects the extent to which members of a group are tightly connected~\cite{granovetter1973strength-weak-ties, moody2003structural-cohesion}. It is a key determinant of group stability, collaboration effectiveness, and information sharing. High cohesion suggests strong inter-member relationships, reinforcing shared understanding and coordination. Low cohesion, on the other hand, indicates fragmentation, where communication and cooperation may be limited, leading to inefficiencies in task execution. By quantifying cohesion, we assess whether a group operates as a unified entity or consists of loosely associated individuals.

Influence captures how control, attention, and decision-making authority are distributed within the group~\cite{bonacich1987power-centrality}. In social and collaborative settings, some individuals naturally emerge as leaders or central figures in discussions, guiding group interactions and shaping overall behavior. The degree of influence varies based on interaction patterns, and an even distribution of influence suggests a more participatory structure, whereas dominance by a few individuals can indicate hierarchical tendencies. Understanding influence allows us to determine whether group behavior is shaped by collective participation or driven by a subset of members.

Connectivity represents the structural resilience of a group’s interaction network~\cite{holme2002edge}. A well-connected group ensures that information flows efficiently between members, reducing dependency on specific individuals for communication. High connectivity enhances adaptability and robustness, as multiple interaction pathways exist, making the group less vulnerable to disruptions. Conversely, low connectivity results in fragile structures where communication breakdowns or the removal of key members can severely impact the group’s ability to function~\cite{freeman1977set}. By characterizing connectivity, we can evaluate the group’s ability to sustain collaboration under changing conditions.

By integrating these three dimensions, we provide a holistic assessment of group dynamics, allowing for interpretations beyond simple pairwise interactions. This characterization enables a structured understanding of group behavior.

\subsection{Threshold Selection and Justifcation}
\label{sec:appx:threshold-selection}
Cohesion is assessed through clustering variability ($\sigma_C$) and eigenvector variability ($\sigma_{x_i}$). Small-world network studies~\cite{watts1998collective-dynamics-cohesion} demonstrate that highly cohesive groups tend to have low clustering variability, while social cohesion theory indicates that fragmented groups exhibit higher variability in influence. Experimental studies on group collaboration confirm that subgroups with stable interactions exhibit lower clustering variability, whereas loosely connected groups show higher dispersion. Cohesion is classified as high when $\sigma_C < 0.05$ and $\sigma_{x_i} < 0.08$, moderate when $\sigma_C < 0.02$ and $\sigma_{x_i} < 0.04$, and low otherwise. Low clustering and eigenvector variability suggest a well-integrated group, whereas higher values indicate fragmentation and uneven participation.

Influence is measured using mean eigenvector centrality ($\mu_{x_i}$), PageRank variability ($\sigma_P$), and mean Katz centrality ($\mu_{katz}$)~\cite{bonacich1987power-centrality}. Eigenvector centrality captures both direct and indirect influence, PageRank variability determines whether the influence is evenly distributed, and Katz centrality accounts for long-range influence, ensuring that indirect leaders are recognized. The threshold for high influence is set at $\mu_{x_i} > 0.47$, $\sigma_P < 0.02$, and $\mu_{katz} > 0.465$, ensuring that key individuals consistently hold influence without a single dominant node. Moderate influence is assigned when $\mu_{x_i} > 0.49$ and $\sigma_P < 0.065$, indicating active but less concentrated influence, while low influence is assigned otherwise. These thresholds align with studies on leadership and network centrality, which highlight the role of variability in identifying balanced versus hierarchical structures.

Connectivity is evaluated through edge connectivity ($\kappa'(G)$) and path length variability ($\sigma_{PL}$)~\cite{holme2002edge}. Edge connectivity measures the minimum number of edges that must be removed to disconnect the network, and path length variability captures how efficiently information flows. High connectivity is assigned when $\kappa'(G) > 2$ and $\sigma_{PL} < 24$, indicating a robust and stable network. Moderate connectivity occurs when $\kappa'(G) > 1$ and $\sigma_{PL} < 50$, suggesting some structural redundancy, while fragile connectivity occurs when these conditions are not met, meaning that a single connection loss could lead to fragmentation. These thresholds are informed by studies on network robustness and efficiency, which emphasize the importance of redundancy and stable communication pathways in group stability.

The thresholds were validated across multiple datasets, including real-world social interaction networks, synthetic networks generated using Erdős–Rényi and Barabási–Albert models, and studies of organizational group structures. These values provide a balance between theoretical rigor and empirical robustness, ensuring that classification remains interpretable across different application domains.

\section{Terminology}
\label{sec:appx:terms}
\subsection{Objective  versus Subjective Accuracy}
\label{sec:appx:accuracy}
Objective accuracy, often used in classification tasks, typically refers to a comparison against a well-defined, verifiable ground truth. In tasks like image classification or object detection, ground truth labels are predefined and universally agreed upon, making objective accuracy a reliable metric. However, in our case, an objective accuracy measure would require predefined, universally accepted labels for each image’s emotional classification, which is impractical given the inherent subjectivity of emotional perception. Emotional responses are influenced by personal experiences, cultural backgrounds, and contextual factors, leading to variations in how individuals interpret and categorize images~\cite{gebhardtDetectingUsersEmotional2024}. Therefore, objective accuracy is not meaningful for this task, as it would fail to capture emotional classification's fluid and interpretative nature. Instead, we rely on subjective accuracy, which measures how well users’ categorizations align with conventional emotional mappings derived from the valence-arousal model. This approach allows us to study the dynamics of consensus-building in collaborative decision-making. By analyzing how groups converge (or diverge) in their classifications, we gain insights into the negotiation processes, social influences, and cognitive alignment required to form shared interpretations. This perspective is particularly important in collaborative MR environments, where achieving consensus among multiple users is essential for effective teamwork and decision-making.

\subsection{Additional Measures Information}
\label{sec:appx:additional-questionarrie}
We designed a group behavior characterization questionnaire to assess key collaboration aspects, including contribution awareness, joint attention, proximity impact, conversational support, and overall collaboration, as summarized in~\ref{table:custom-questionnaire}. These questions capture participant perceptions of engagement, shared focus, and communication, drawing from prior research on social awareness, joint attention, proximity effects, and group cohesion. This questionnaire provides subjective insights that complement sensor-based analysis, offering a holistic understanding of group interactions in MR environments.
\begin{table}[t]
\centering
\caption{Proposed Group Behavior Characterization Questionnaire.}
\vspace{-0.3cm}
\label{table:custom-questionnaire}
\resizebox{\textwidth}{!}{
\begin{tabular}{|c|l|l|}
\toprule \hline
             & \makecell[c]{\textbf{Proposed Question}} & \makecell[c]{\textbf{Description}} \\ \hline  \hline
    
        \makecell[c]{Contribution \\Awareness} & \makecell[c]{How much did you feel other group  members\\ contributed during the task?} & \makecell[c]{
        Assesses participants' perceptions of their peers' contributions. Inspired\\ by the work on social awareness and collaboration by~\citet{carroll2006awareness-teamwork}. } \\ \hline
        
        \makecell[c]{Joint Attention\\ Awareness} & \makecell[c]{How often did you feel you were paying attention\\ to the same virtual object as other participants?} & \makecell[c]{To evaluate shared focus, which is vital for effective collaboration. \\Based on research on joint attention by~\citet{tomasello2014joint-attention}.} \\ \hline

        \makecell[c]{Proximity \\Impact} & \makecell[c]{I felt that my proximity to others affected\\ my collaboration during the task.} & \makecell[c]{Examines how physical closeness influences communication\\ and group performance. Inspired by the work on\\ proximity frameworks by~\citet{gonzalez2015framework-proximity}.} \\ \hline
        
        \makecell[c]{Conversational \\Support} & \makecell[c]{How much did the group conversation help you\\ understand the task and contribute effectively?} & \makecell[c]{Assesses the role of dialogue in shaping understanding\\ and participation. Based on the conversational analysis\\ in group dynamics by~\citet{goodwin1990conversation-analysis}.
        } \\ \hline
        
        \makecell[c]{Group \\Collabration} & \makecell[c]{The group worked together effectively\\ to complete the task.} & \makecell[c]{Evaluates overall group collabortaion and unity. Based on\\  research on group performance and cohesion by~\citet{forsyth2021-group-cohesion}} \\ \hline
        \bottomrule
\end{tabular}
}
\end{table}

\section{Pilot Study}\label{app:pilot}
To refine the experimental design and optimize the main user study, we conducted a pilot study with one group of four participants. The primary objectives of this pilot study were to assess the feasibility and clarity of the setup, test device functionality, and evaluate the participants' interactions with the virtual objects. Observations during the pilot study highlighted several key insights that informed adjustments to the main study protocol.

Firstly, participants spent a significant amount of time becoming accustomed to interacting with virtual objects. To streamline this, a tutorial was added, featuring a simplified task with two images and labels to help users familiarize themselves with the controls. Additionally, we observed that enabling distance grabbing for virtual objects allowed participants to remain stationary, which reduced spatial interaction among group members. To address this, distance grabbing was disabled, encouraging participants to physically move closer to objects, thereby promoting more realistic interactions that are similar to real-world collaborative settings.
The pilot study also provided insights into the nature of conversational interactions. Notably, the conversation was rarely directed toward specific individuals. This observation aligned with previous research suggesting a focus on virtual objects over co-participants in MR environments~\cite{prytzImportanceEyecontactCollaboration2010} and justified our assumption of group-directed conversation for data analysis.
Based on these insights, we adjusted the main study protocol to enhance task clarity, foster more natural group behaviors, and provide a more reliable setup for capturing group behavior in the MR environment.

\section{Additional Experimental Results}
\label{sec:appx:additional-results}

\subsection{Additional Sociograms and Task Level Measures}\label{appendix:sociograms}
\begin{figure}[h]
  \centering
  \includegraphics[width=0.9\linewidth]{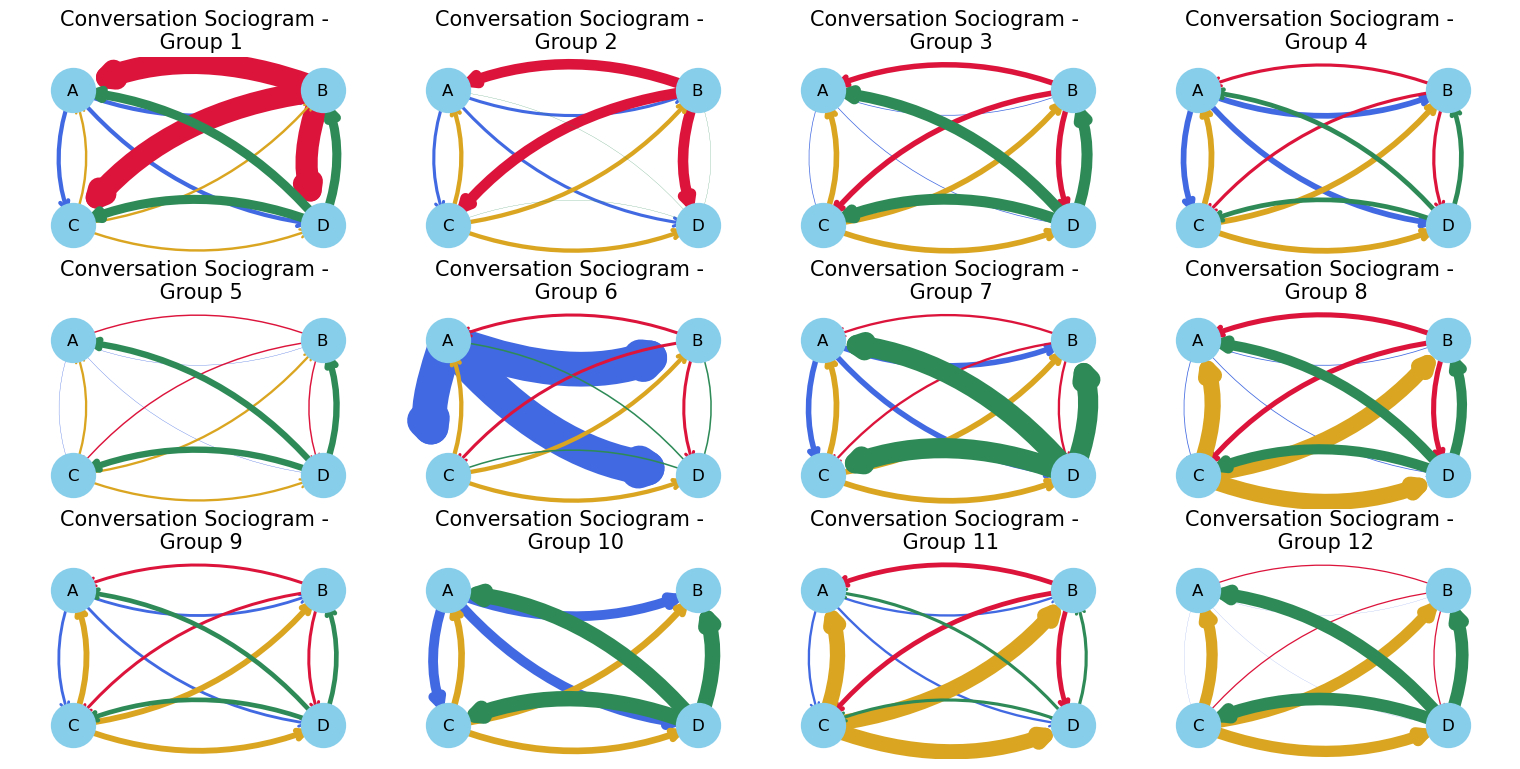}
  \caption{Sociograms representing conversation duration across all groups. The edge thickness reflects the total duration of conversations between participants, illustrating the intensity and frequency of interactions.}
  \label{fig:conversation-sociogram}
\end{figure}

\begin{figure}[h] 
    \centering \includegraphics[width=0.9\linewidth]{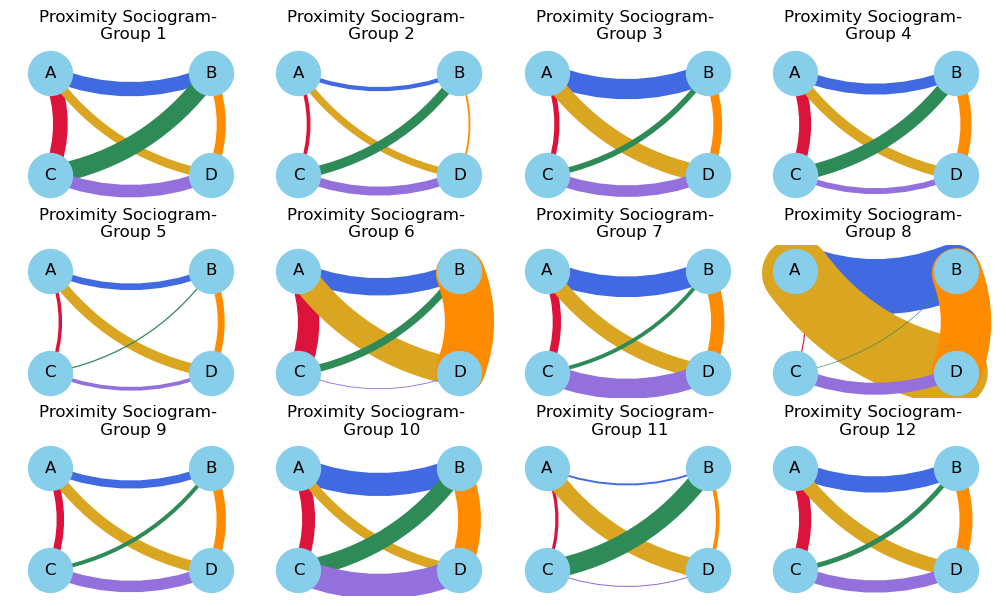} \caption{Sociograms representing proximity duration across all groups. The The edge thickness represents the total duration of time participants spent within a defined proximity threshold, reflecting the frequency and intensity of spatial interactions.} 
    \label{fig:proximity-sociogram} 
\end{figure}

\begin{figure}[h]
    \centering \includegraphics[width=0.9\linewidth]{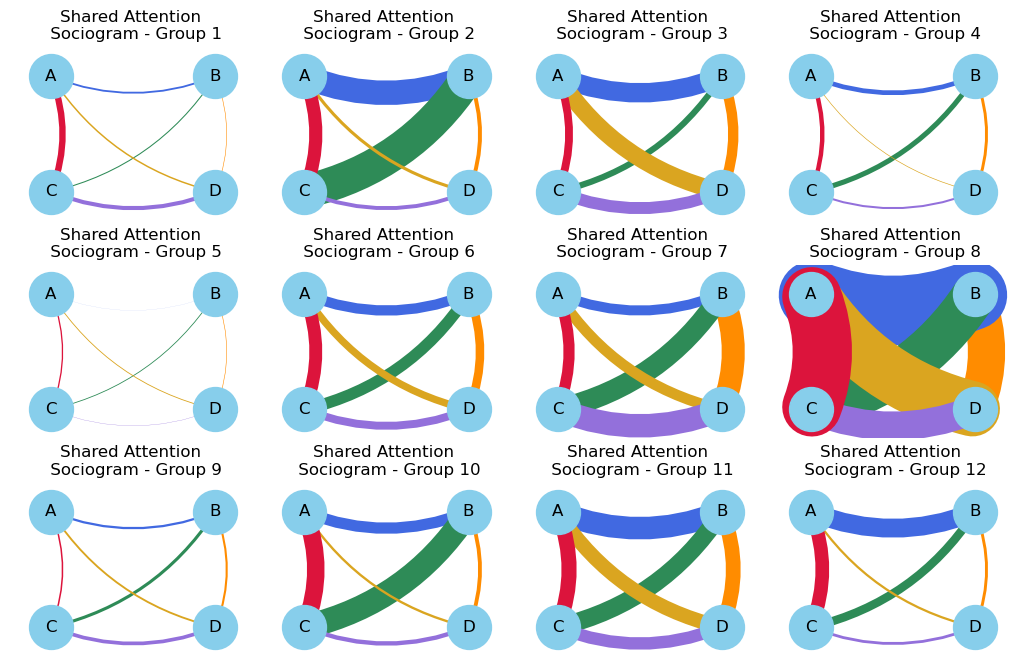} \caption{Sociograms representing shared attention duration across all groups. The edge thickness represents the cumulative duration of shared gaze on the same virtual objects, highlighting the extent of joint focus and collaborative attention among participants.} 
    \label{fig:sharedatt-sociogram} 
\end{figure}

\begin{figure}[h]
  \centering
  \includegraphics[width=0.9\linewidth]{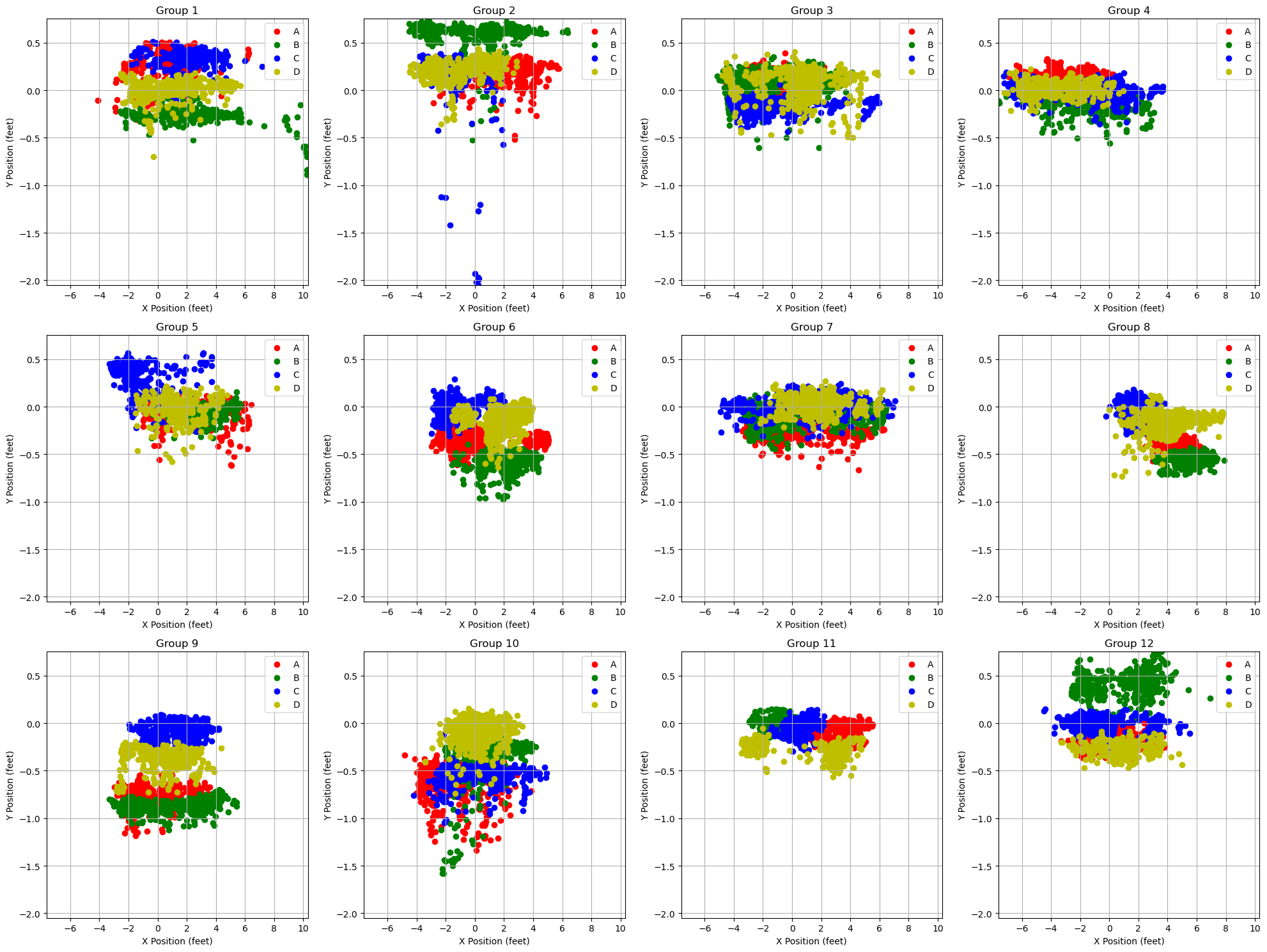}
  \caption{Location heatmap of each participant in different groups.
  }
  \label{fig:location-heatmap-appendix}
\end{figure}

\begin{figure}[ht]
  \centering
  \begin{subfigure}{=\textwidth}  
    \centering
    \includegraphics[width=\linewidth]{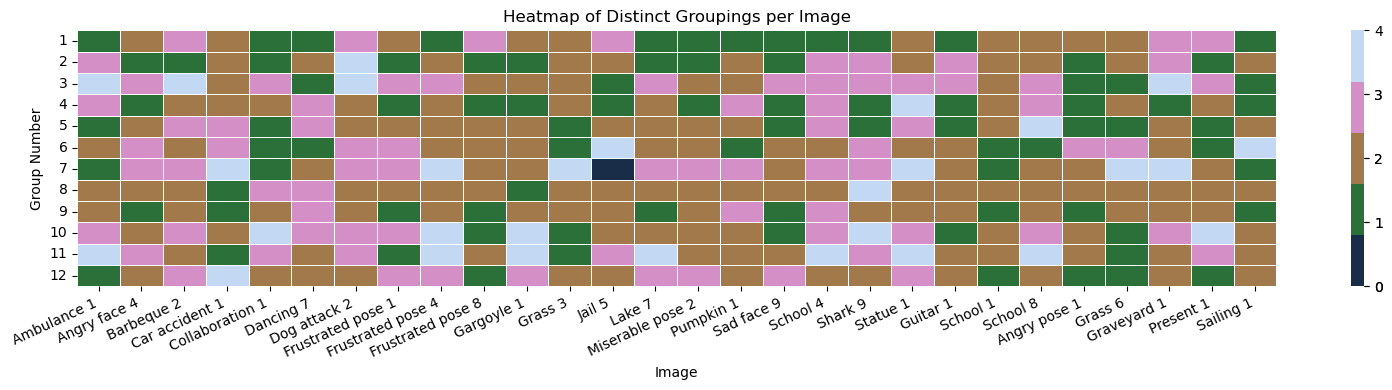}
    \caption{}
    \label{fig:distinct-grouping}
  \end{subfigure}
  \hspace{0.05\textwidth} 
  \begin{subfigure}{\textwidth}  
    \centering
    \includegraphics[width=\linewidth]{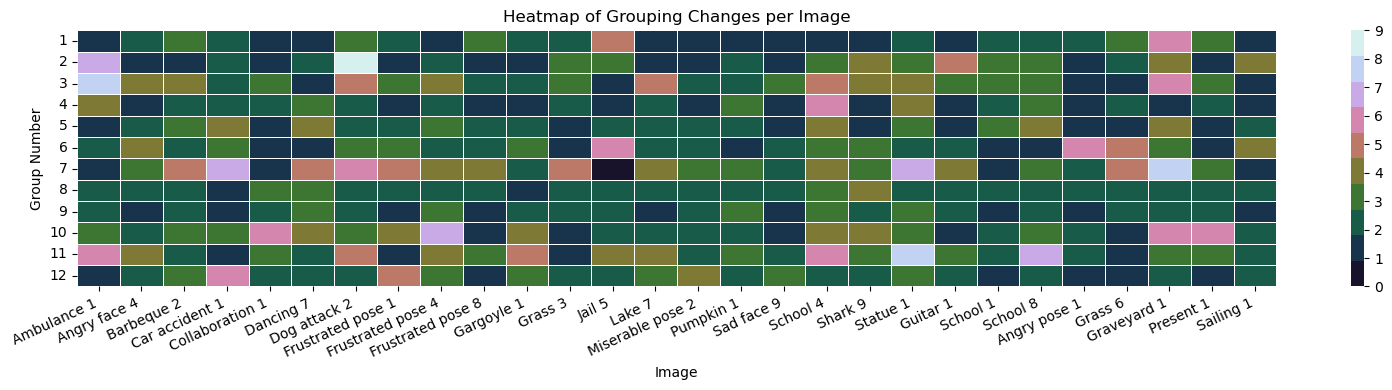}
    \caption{}
    \label{fig:grouping-changes}
  \end{subfigure}
  \caption{(a) Distinct labels for each image recorded for a group. (b) Total label changes for each image recorded for a group. Example: A moved Ambulance to Tense, B moved Ambulance to Bored, and then A moved Ambulance back to Tense; this counts image label changes as $3$ and distinct image labels as $2$. 
  }
  \label{fig:grouping-interactions}
\end{figure}

\begin{figure}
  \centering
  \includegraphics[width=\linewidth]{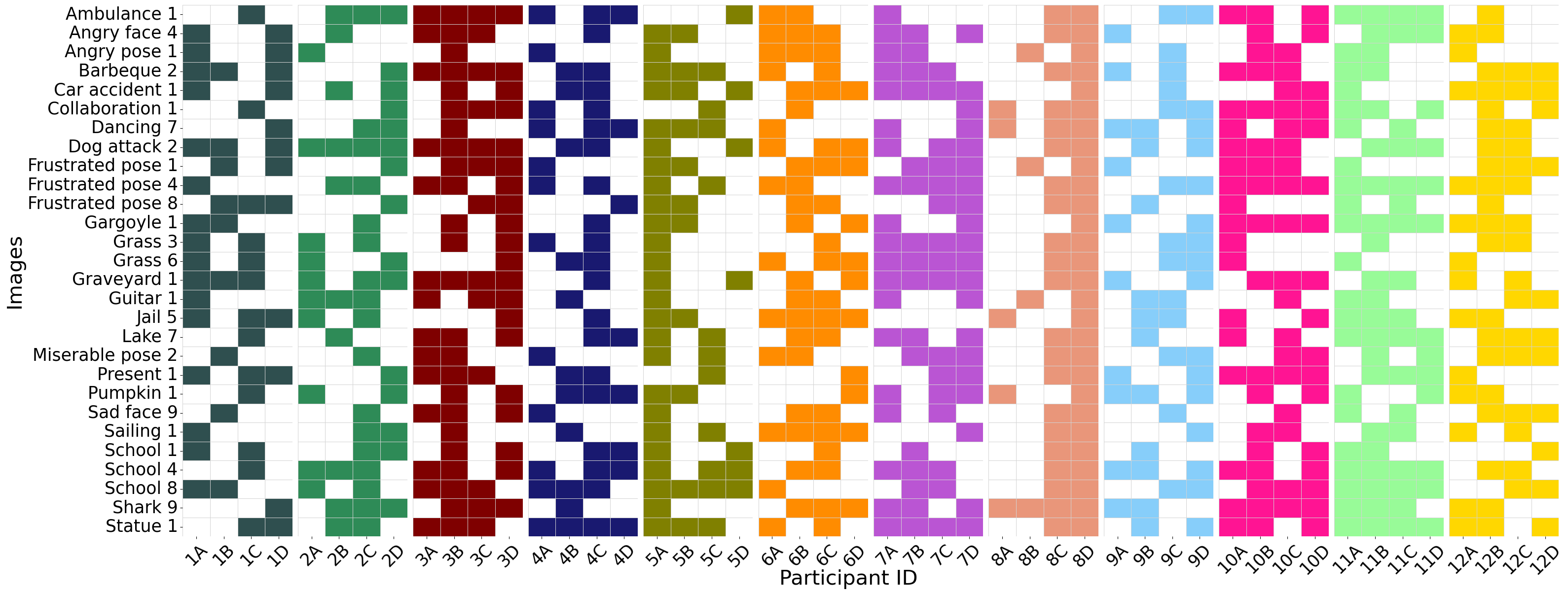}
  \caption{Heatmap representing which users interacted with each image.}
  \label{fig:user-list-image}
\end{figure}

\subsection{Analysis of Mapping between Graph Attributes and Group Characteristics}
\label{sec:sna-analyis}
\autoref{tab:group-characteristics-sociogram}, presents graph attributes mapping to group characteristics across sociograms. The conversation sociogram shows varied cohesion, with some groups (e.g., Groups 3 and 4) being highly cohesive, while others (e.g., Groups 1, 2, and 5) appear more fragmented. The proximity sociogram indicates that cohesive groups (e.g., Groups 1, 2, and 9) maintain tight-knit clustering, reflecting stronger physical interactions, while fragmented groups show loose-knit structures. The shared attention sociogram highlights resilient connectivity in Groups 1, 4, and 10, contrasting with other groups' more mixed structures. This distribution reveals that attention-based interactions contribute differently to group dynamics than conversation and proximity.

\begin{table}[t]
\centering
\scriptsize
\caption{Group Characteristics Scores Based on Sociograms. }

\vspace{-0.35cm}
\label{tab:group-characteristics-sociogram}
\begin{tabular}{|c|l|l|l|l|}
\hline
\multicolumn{1}{|l|}{Group ID} & Sociogram    & {\color[HTML]{330001} Cohesion} & Influence                                & Connectivity                    \\ \hline
                               & Conversation & low (1.0)                       & high (0.33), moderate (0.33), low (0.33) & resilient (0.5), fragile (0.5)  \\ \cline{2-5} 
                               & Proximity    & moderate (1.0)                  & high (0.33), moderate (0.33), low (0.33) & resilient (0.5), fragile (0.5)  \\ \cline{2-5} 
\multirow{-3}{*}{Group 1}      & Attention    & moderate (0.5), low (0.5)       & moderate (0.33), high (0.33), low (0.33) & resilient (1.0)                 \\ \hline
                               & Conversation & low (1.0)                       & high (0.33), moderate (0.33), low (0.33) & resilient (0.5), fragile (0.5)  \\ \cline{2-5} 
                               & Proximity    & moderate (1.0)                  & high (0.33), moderate (0.33), low (0.33) & resilient (0.5), moderate (0.5) \\ \cline{2-5} 
\multirow{-3}{*}{Group 2}      & Attention    & low (1.0)                       & low (0.67), high (0.33)                  & resilient (1.0)                 \\ \hline
                               & Conversation & moderate (1.0)                  & high (0.33), moderate (0.33), low (0.33) & resilient (0.5), fragile (0.5)  \\ \cline{2-5} 
                               & Proximity    & moderate (0.5), low (0.5)       & high (0.33), moderate (0.33), low (0.33) & resilient (0.5), fragile (0.5)  \\ \cline{2-5} 
\multirow{-3}{*}{Group 3}      & Attention    & moderate (1.0)                  & high (0.33), moderate (0.33), low (0.33) & resilient (0.5), moderate (0.5) \\ \hline
                               & Conversation & moderate (0.5), high (0.5)      & high (0.67), low (0.33)                  & resilient (0.5), moderate (0.5) \\ \cline{2-5} 
                               & Proximity    & moderate (0.5), high (0.5)      & high (0.67), moderate (0.33)             & resilient (0.5), fragile (0.5)  \\ \cline{2-5} 
\multirow{-3}{*}{Group 4}      & Attention    & low (1.0)                       & moderate (0.67), low (0.33)              & resilient (1.0)                 \\ \hline
                               & Conversation & low (1.0)                       & high (0.67), low (0.33)                  & resilient (0.5), moderate (0.5) \\ \cline{2-5} 
                               & Proximity    & low (1.0)                       & moderate (0.33), high (0.33), low (0.33) & resilient (0.5), moderate (0.5) \\ \cline{2-5} 
\multirow{-3}{*}{Group 5}      & Attention    & moderate (0.5), low (0.5)       & high (0.67), moderate (0.33)             & resilient (1.0)                 \\ \hline
                               & Conversation & low (1.0)                       & moderate (0.33), high (0.33), low (0.33) & resilient (0.5), fragile (0.5)  \\ \cline{2-5} 
                               & Proximity    & low (1.0)                       & moderate (0.67), low (0.33)              & resilient (0.5), fragile (0.5)  \\ \cline{2-5} 
\multirow{-3}{*}{Group 6}      & Attention    & moderate (1.0)                  & high (0.33), moderate (0.33), low (0.33) & resilient (0.5), moderate (0.5) \\ \hline
                               & Conversation & low (1.0)                       & high (0.33), moderate (0.33), low (0.33) & resilient (0.5), fragile (0.5)  \\ \cline{2-5} 
                               & Proximity    & moderate (1.0)                  & high (0.33), moderate (0.33), low (0.33) & resilient (0.5), fragile (0.5)  \\ \cline{2-5} 
\multirow{-3}{*}{Group 7}      & Attention    & low (1.0)                       & high (0.33), moderate (0.33), low (0.33) & resilient (0.5), moderate (0.5) \\ \hline
                               & Conversation & low (1.0)                       & high (0.33), moderate (0.33), low (0.33) & resilient (0.5), fragile (0.5)  \\ \cline{2-5} 
                               & Proximity    & low (1.0)                       & low (0.67), high (0.33)                  & resilient (0.5), fragile (0.5)  \\ \cline{2-5} 
\multirow{-3}{*}{Group 8}      & Attention    & moderate (1.0)                  & high (0.33), moderate (0.33), low (0.33) & resilient (0.5), fragile (0.5)  \\ \hline
                               & Conversation & moderate (0.5), high (0.5)      & high (0.67), moderate (0.33)             & resilient (0.5), moderate (0.5) \\ \cline{2-5} 
                               & Proximity    & moderate (1.0)                  & high (0.67), moderate (0.33)             & resilient (0.5), fragile (0.5)  \\ \cline{2-5} 
\multirow{-3}{*}{Group 9}      & Attention    & moderate (1.0)                  & high (0.67), moderate (0.33)             & resilient (1.0)                 \\ \hline
                               & Conversation & low (1.0)                       & high (0.33), moderate (0.33), low (0.33) & resilient (0.5), fragile (0.5)  \\ \cline{2-5} 
                               & Proximity    & moderate (1.0)                  & high (0.33), moderate (0.33), low (0.33) & resilient (0.5), fragile (0.5)  \\ \cline{2-5} 
\multirow{-3}{*}{Group 10}     & Attention    & low (1.0)                       & low (0.67), high (0.33)                  & resilient (1.0)                 \\ \hline
                               & Conversation & moderate (0.5), low (0.5)       & high (0.33), moderate (0.33), low (0.33) & resilient (0.5), fragile (0.5)  \\ \cline{2-5} 
                               & Proximity    & high (1.0)                      & high (0.67), low (0.33)                  & resilient (1.0)                 \\ \cline{2-5} 
\multirow{-3}{*}{Group 11}     & Attention    & moderate (1.0)                  & high (0.67), moderate (0.33)             & resilient (0.5), moderate (0.5) \\ \hline
                               & Conversation & low (1.0)                       & moderate (0.33), high (0.33), low (0.33) & resilient (0.5), fragile (0.5)  \\ \cline{2-5} 
                               & Proximity    & moderate (1.0)                  & high (0.33), moderate (0.33), low (0.33) & resilient (0.5), fragile (0.5)  \\ \cline{2-5} 
\multirow{-3}{*}{Group 12}     & Attention    & low (1.0)                       & low (0.67), high (0.33)                  & resilient (1.0)                 \\ \hline
\end{tabular}
\end{table}

\subsection{Analysis of Key Graph Attributes}
\label{sec:appx:key-graph-att:analysis}
\autoref{tab:chi-squared} presents the analysis of relationships between key graph attributes using chi-square tests for categorical interactions and one-way ANOVA for continuous variables. The goal is to evaluate whether cohesion, influence, clustering, connectivity, and centralization are significantly associated with conversation, proximity and shared attention sociograms. Although the primary focus is cohesion, influence, and connectivity, the chi-square and ANOVA results also analyze relationships with centralization and clustering. These attributes provide additional insights into how groups self-organize, revealing whether certain structural characteristics emerge naturally from interaction patterns. Centralization helps assess whether group leadership or dominance affects cohesion and influence. Clustering provides insight into subgroup formation and localized group behaviors that may impact overall connectivity and influence distribution.

By examining these broader relationships, we can better understand how different structural aspects shape group coordination and decision-making processes within each sociogram type. \autoref{tab:chi-squared} presents the results of chi-square tests for categorical relationships between graph attributes and one-way ANOVA for continuous variables across different sociograms. The goal is to evaluate whether clustering, connectivity, and centralization are significantly associated within conversation, proximity, and shared attention sociograms.

In the conversation sociogram, chi-square tests showed no significant associations between clustering, connectivity, or centralization. ANOVA results indicated no significant differences in cohesion across connectivity types ($F = 3.73$, $p = 0.085$) or influence across clustering types ($F = 0.23$, $p = 0.637$). This suggests that in conversation-based interactions, these structural attributes operate independently rather than forming strong dependencies.
For the proximity sociogram, clustering and connectivity showed no significant relationship ($\chi^2 = 0.91$, $p = 0.632$), but clustering and centralization had a near-significant association ($\chi^2 = 2.80$, $p = 0.093$). This indicates a potential tendency for more centralized groups to exhibit distinct clustering patterns. ANOVA results showed no significant effect of connectivity or clustering on cohesion ($F = 0.28$, $p = 0.757$) or influence ($F = 3.15$, $p = 0.109$), suggesting that spatial proximity alone does not strongly impact cohesion or influence distribution.
In the shared attention sociogram, clustering and centralization were significantly associated ($\chi^2 = 3.93$, $p = 0.047$), indicating that groups with high clustering are more likely to exhibit centralized structures. ANOVA revealed significant differences in influence across clustering types ($F = 5.25$, $p = 0.047$), implying that higher clustering in shared attention leads to higher influence distribution. However, cohesion across connectivity types was not significant ($F = 2.89$, $p = 0.113$), meaning that connectivity strength does not necessarily lead to greater cohesion in shared attention scenarios.

\begin{table}[ht]
\centering
\scriptsize
\caption{Chi-Square and ANOVA Results across Contexts.}
\label{tab:chi-squared}
\begin{tabular}{|c|c|c|c|}
\hline
 \textbf{Sociogram}  & Test & Test Statistic & p-value \\ \hline
 \multirow{5}{*}{Conversation}  & Clustering vs Connectivity & 0.3315 & 0.5648\\ \cline{2-4}
 & Clustering vs Centralization & 0.0000 & 1.0000 \\ \cline{2-4}
& Connectivity vs Centralization & 0.0000 & 1.0000 \\ \cline{2-4}
& Cohesion across Connectivity (ANOVA) & 0.0032 & 0.9559\\ \cline{2-4}
& Influence across Clustering (ANOVA) & 0.0698 & 0.7976 \\ \hline

 \multirow{5}{*}{Proximity} & Clustering vs Connectivity & 0.9167 & 0.6323 \\ \cline{2-4}
& Clustering vs Centralization & 2.8073 & 0.0938 \\ \cline{2-4}
& Connectivity vs Centralization & 1.7569 & 0.4154\\ \cline{2-4}
& Cohesion across Connectivity (ANOVA) & 1.6926 & 0.2438 \\ \cline{2-4}
& Influence across Clustering (ANOVA) & 28.4026 & 0.0005 \\ \hline

 \multirow{5}{*}{Attention} & Clustering vs Connectivity & 1.2768 & 0.5281 \\ \cline{2-4}
& Clustering vs Centralization & 3.9327 & 0.0474 \\ \cline{2-4}
& Connectivity vs Centralization & 3.4375 & 0.1793 \\ \cline{2-4}
& Cohesion across Connectivity (ANOVA) & 1.6268 & 0.2554 \\ \cline{2-4}
& Influence across Clustering (ANOVA) & 2.0140 & 0.1895 \\ \hline
\end{tabular}
\end{table}

In the proximity sociogram, clustering and connectivity showed no significant association ($\chi^2 = 0.91$, $p = 0.632$), but clustering and centralization had a near-significant relationship ($\chi^2 = 2.80$, $p = 0.093$). ANOVA showed no significant impact of connectivity and clustering on cohesion ($F = 0.28$, $p = 0.757$) or influence ($F = 3.15$, $p = 0.109$).
In the shared attention sociogram, clustering and centralization were significantly associated ($\chi^2 = 3.93$, $p = 0.047$). ANOVA indicated significant differences in influence across clustering types ($F = 5.25$, $p = 0.047$), while cohesion across connectivity types was not significant ($F = 2.89$, $p = 0.113$). This implies that clustering in shared attention impacts influence but not cohesion. These analyses underline the relationships between graph attributes and sociograms, suggesting that interaction types shape group characteristics.

\subsection{Correlation Analysis of Task and Individual Behavior Metrics with Group Behavior}
\label{sec:appx:correlation-analysis}
The results in ~\autoref{tab:correlation-task-individual-vs-group} show the correlation analysis between task/individual metrics and group behavior. Shapiro-Wilk tests assessed normality, revealing non-normal distributions ($p-values < 0.05$). Thus, Spearman's correlation was used for robust non-parametric analysis~\cite{hauke2011comparison-spearman-vs-pearson}.  Metrics such as the number of images grabbed, image grouping changes, and task completion time showed weak correlations with group behavior (e.g., images grabbed: $0.06, p = 0.86$). Image grouping decisions had a slightly higher, non-significant correlation ($0.23, p = 0.4790$). Presence showed a moderate negative correlation with group state ($-0.47, p = 0.136$), while collaboration had the strongest, though non-significant, correlation ($-0.53, p = 0.086$). Other metrics like attention had negligible correlations (e.g., $-0.09, p = 0.790$).
The results show minimal correlation between task performance metrics and group behavior, indicating that group cohesiveness or fragmentation has little impact on objective task performance. However, presence and collaboration metrics suggest that cohesive groups may experience stronger engagement and teamwork. Although p-values for these metrics are insignificant, trends align with prior research, which underscores the potential importance of subjective experience metrics, such as presence and collaboration, in understanding group behaviors. Future studies with larger samples could further explore these observations.
\label{sec:appx:task-behavior-metric-correlation}
\begin{table}[!t]
    \centering
    \caption{Correlation of Task and Individual Behavior Metrics with Group Behavior.}
    \small
    \label{tab:correlation-task-individual-vs-group}
    \resizebox{\textwidth}{!}{
    \begin{tabular}{|c|l|c|c|c|c|}
        \toprule
        \hline
        & \textbf{Metric} & \textbf{Target Normality (p)} & \textbf{Metric Normality (p)} & \textbf{Spearman Correlation  Coefficient} & \textbf{P-value} \\ 
        \hline
        \hline
        \multirow{9}{*}{\rotatebox{90}{\textbf{Task and Performance}}} &
        Images Grabbed & 0.0001 & 0.0199 & 0.0600 & 0.8608 \\ 
        & Total Image Grabbing & 0.0001 & 0.6663 & -0.1497 & 0.6603 \\ 
        & Image Grouping Decision & 0.0001 & 0.0000 & 0.2390 & 0.4790 \\ 
        & Image Grouping Overridden & 0.0001 & 0.0191 & 0.0000 & 1.0000 \\ 
        & Images Looked At & 0.0001 & 0.0048 & 0.0000 & 1.0000 \\ 
        & Distinct Groupings & 0.0001 & 0.0199 & 0.0600 & 0.8608 \\ 
        & Grouping Changes & 0.0001 & 0.2744 & -0.1195 & 0.7263 \\ 
        & Completion Time (seconds) & 0.0001 & 0.4953 & -0.1195 & 0.7263 \\ 
        & Subjective Accuracy (\%) & 0.0001 & 0.1904 & -0.1504 & 0.6589 \\ \hline
        \multirow{7}{*}{\rotatebox{90}{\textbf{Behavior}}} &
        Presence Score & 0.0001 & 0.5625 & -0.4781 & 0.1369 \\ \cline{2-6}
        & TLX Score & 0.0001 & 0.4707 & 0.0898 & 0.7928 \\ \cline{2-6}
        & \hspace{1em}Cohesion & 0.0001 & 0.1358 & -0.2151 & 0.5253 \\ 
        & \hspace{1em}Attention & 0.0001 & 0.2919 & -0.0907 & 0.7909 \\ 
        & \hspace{1em}Proximity & 0.0001 & 0.9033 & -0.0605 & 0.8599 \\ 
        & \hspace{1em}Conversation & 0.0001 & 0.5016 & 0.0613 & 0.8579 \\
        & \hspace{1em}Collaboration & 0.0001 & 0.0124 & -0.5396 & 0.0867 \\ 
        \hline
        \bottomrule
    \end{tabular}
    }
\end{table}

\begin{table}[t]
    \centering
    \caption{Effect of Group Behavior on Task and Individual Behavior Metrics.}
    \vspace{-0.3cm}
    \small
    \label{tab:task-behvior-vs-group-label}
    \resizebox{\textwidth}{!}{
    \begin{tabular}{|c||l||c|c|c||c|c|c||c|c|}
    \toprule
    \hline
    &
    \textbf{Metric} & \textbf{\makecell[c]{Shapiro \\Cohesive (p)}}  & \textbf{\makecell[c]{Shapiro \\ Fragmented (p)}} & 
    \textbf{\makecell[c]{Levene \\ Test (p)}} & 
    \textbf{\makecell[c]{T-test \\ Statistic}} & 
    \textbf{\makecell[c]{T-test \\ p-value}} & 
    \textbf{\makecell[c]{Mann-Whitney \\ U test}} & 
    \textbf{\makecell[c]{Cohen d}} &
    \textbf{\makecell[c]{ Power}} \\ \hline
    \hline
    \multirow{9}{*}{\rotatebox{90}{\textbf{Task and Performance}}} &
    Images Grabbed & 0.0582 & 0.2842 & 0.1397 & 0.2792 & 0.7864 & N/A & 0.18 & 0.06 \\ 
    & Total Image Grabbing & 0.5958 & 0.8494 & 0.2468 & -0.3683 & 0.7212 & N/A & -0.23 & 0.06 \\ 
    & Image Grouping Decision & 1.0000 & 0.0000 & 0.4790 & 0.7385 & 0.4790 & N/A & 0.53& 0.12\\ 
    & Image Grouping Overridden & 0.0582 & 0.3207 & 0.1698 & 0.2524 & 0.8064 & N/A & 0.16 & 0.06 \\ 
    & Images Looked At & 0.2725 & 0.0219 & 0.8722 & 0.1505 & 0.8837 & N/A & 0.10 & 0.05\\ 
    & Distinct Groupings & 0.0582 & 0.2842 & 0.1397 & 0.2792 & 0.7864& N/A & 0.18 & 0.06 \\ 
    & Grouping Changes & 0.1190 & 0.3415 & 0.2294 & -0.0223 & 0.9827& N/A & -0.01 & 0.05 \\
    & Completion Time (seconds) & 0.7055 & 0.6321 & 0.3753 & -0.3410 & 0.7409 & N/A & -0.21 & 0.06 \\ 
    & Accuracy (\%) & 0.2616 & 0.2682 & 0.5891 & 0.0322 & 0.9750&N/A & 0.02 & 0.05 \\ 
    \hline
    \hline
    \multirow{7}{*}{\rotatebox{90}{\textbf{Behavior}}} &
    Presence Score & $0.9231$ & $0.7001$ & $0.1129$ & $-1.8731$ & 0.0938& N/A & -1.17 & 0.39 \\ \cline{2-10}
    & TLX Score & 0.9902 & 0.6381 & 0.8021 & 0.5093 & 0.6228& N/A & 0.32 & 0.07\\ \cline{2-10}
    
    &  \hspace{1em}Cohesion & 0.8500 & 0.1343 & 0.8141 & -0.5884 &  0.5707& N/A & -0.37 & 0.08 \\ 
    &  \hspace{1em}Attention & 0.3690 & 0.3301 & 0.7444 & -0.6818 & 0.5125 & N/A & -0.43 & 0.09 \\ 
    &  \hspace{1em} Proximity & 0.7966 & 0.6037 & 0.7201 & -0.4782 & 0.6439& N/A & -0.30 & 0.07 \\ 
    &  \hspace{1em}Conversation & 0.5719 & 0.1953 & 0.7416 & 0.4984 & 0.6301& N/A & 0.31 & 0.07 \\ 
    &  \hspace{1em}Collaboration & 0.2725 & 0.0261 & 0.1722 & -1.8608 & 0.0957&5.5 & N/A & N/A \\ 
    \hline
    \bottomrule
\end{tabular}
}
\end{table}
\subsection{Group Behavior Impact on Task Performance and Behavioral Metrics}
\label{sec:task-behavior-metrics-vs-group}
This section outlines the statistical analysis assessing the impact of group behavior on task performance and individual behavior metrics. ~\autoref{tab:task-behvior-vs-group-label} summarizes the results. The Shapiro-Wilk test~\cite{shapiro1965analysis} confirms normality for most metrics, allowing parametric analysis, with exceptions like image grouping and collaboration in fragmented groups ($p = 0.0261$). Levene’s test~\cite{schultz1985levene} shows homogeneity of variances (p > 0.05), supporting the use of independent sample t-tests~\cite{kim2015t-test}. When normality is violated, the Mann-Whitney U test~\cite{mcknight2010mann-whitney} is used.
Cohen’s d~\cite{cohen2013statistical-cohen-d-power-analysis} shows small effect sizes, suggesting minimal practical differences between groups aligned with low power values. Although task performance metrics show little difference, presence and collaboration metrics hint at more variation. The presence score approaches significance with a medium effect size, indicating cohesive groups may feel more engaged. Group collaboration shows distribution differences but lacks strict significance. Overall, subjective metrics highlight the benefits of group cohesion in perceived engagement and interaction, even if task performance remains similar. 

\subsection{Additional Behavioral Measures}
\label{sec:appx:behavioral-measure}
\autoref{tab:survey-summary-stats} summarizes descriptive statistics, showing participants' subjective experiences and response variability. Metrics like PQ ($\mu 5.46$) and IPQ ($\mu 4.39$) indicate moderate presence levels. The PQ-REAL subscale has a mean of 5.16 ($\sigma 1.1$), suggesting moderate realism, while the ACT subscale shows a high mean of 5.8, indicating strong perceived action capability. INV subscale variability ($\mu 3.68, \sigma 1.53$) highlights differing engagement levels. SP ($\mu=5.07$) reflects strong spatial awareness. The mean NASA TLX score of 2.3 points to a low perceived workload. Group cohesion and collaboration scored high ($\mu 6.55$ and $6.64$), while group proximity varied ($\mu=3.93$, higher SD), indicating differing closeness perceptions.
\begin{table}[]
\caption{Summary of descriptive and statistical results for PQ and IPQ questionnaire and their subscales, NASA TLX scores, and our custom group behavior survey. 
We provide the following statistical metrics: mean ($\mu$) to indicate the average response, standard deviation ($\sigma$) to measure the response variability, standard error of the mean to estimate the precision of the mean, 95\% confidence interval (CI) for the mean to provide a range within which the true population mean likely lies, the 5th and 95th percentiles (CI[5\%] and CI[95\%]) to highlight the distribution boundaries, and minimum (Min) and maximum (Max) values to show the range of responses. The sub-scales for PQ and IPQ questionnaires are realism (PQ-REAL, IPQ-REAL), possibility to act (ACT), interface quality (IFQUAL), possibility to examine (EXAM), self-evaluation of performance (EVAL), involvement (INV), general presence (GP), and spatial presence (SP).}
\vspace{-0.3cm}
\label{tab:survey-summary-stats}
\scriptsize
\begin{tabular}{|c|c|c|c|c|c|c|c|c|c|}
\hline
\hline
\textbf{Quest.} & \textbf{N} & \textbf{$\mu$} & \textbf{$\sigma$} & \textbf{Std. Error} & \textbf{95\% CI (Mean)} & \textbf{CI[5\%]} &  \textbf{CI[95\%] }& \textbf{Min} & \textbf{Max} \\
\hline
\hline
PQ-REAL & 48 & 5.16 & 1.1 & 0.17 & 4.82 to 5.49 & 3.49 & 6.86 & 2.86 & 7 \\
ACT & 48 & 5.8 & 0.89 & 0.13 & 5.53 to 6.07 & 4.5 & 7 & 3.5 & 7 \\
IFQUAL & 48 & 5.29 & 1.11 & 0.17 & 4.95 to 5.63 & 3.38 & 6.67 & 2.33 & 7 \\
EXAM & 48 & 5.67 & 0.83 & 0.13 & 5.42 to 5.93 & 4.33 & 6.95 & 4 & 7 \\
EVAL & 48 & 5.81 & 0.92 & 0.14 & 5.53 to 6.09 & 4.5 & 7 & 3 & 7 \\
\hline
\hline
\textbf{PQ} & 48 & 5.46 & 0.75 & 0.11 & 5.23 to 5.69 & 4.27 & 6.62 & 4.11 & 6.95 \\
\hline
\hline
INV & 48 & 3.68 & 1.53 & 0.23 & 3.22 to 4.15 & 1.54 & 6.6 & 1 & 7 \\
SP & 48 & 5.07 & 1.21 & 0.18 & 4.71 to 5.44 & 2.89 & 6.77 & 1 & 7 \\
GP & 48 & 5.32 & 1.55 & 0.23 & 4.85 to 5.79 & 2 & 7 & 1 & 7 \\
IPQ-REAL & 48 & 4 & 1.21 & 0.18 & 3.63 to 4.37 & 2.25 & 5.96 & 1.25 & 6.5 \\
\hline
\hline
\textbf{IPQ} & 48 & 4.39 & 1.12 & 0.17 & 4.05 to 4.73 & 2.52 & 6.21 & 1.07 & 6.5 \\
\hline
\hline
\textbf{Quest (PQ+IPQ)} & 48 & 4.92 & 0.86 & 0.13 & 4.66 to 5.18 & 3.73 & 6.17 & 2.69 & 6.72 \\
\hline
\hline
\textbf{NASA TLX} & 48 & 2.3 & 0.93 & 0.14 & 2.01 to 2.58 & 1.03 & 3.8 & 1 & 4.2 \\
\hline \hline
Group Cohesion & 48 & 6.55 & 0.85 & 0.13 & 6.29 to 6.8 & 5 & 7 & 3 & 7 \\
Group Attention & 48 & 5.16 & 1.36 & 0.21 & 4.74 to 5.57 & 3.15 & 7 & 2 & 7 \\
Group Proximity & 48 & 3.93 & 1.99 & 0.3 & 3.33 to 4.54 & 1 & 7 & 1 & 7 \\
Group Conversation & 48 & 6.2 & 1 & 0.15 & 5.9 to 6.51 & 4.15 & 7 & 3 & 7 \\
Group Collaboration & 48 & 6.64 & 0.84 & 0.13 & 6.38 to 6.89 & 5.15 & 7 & 3 & 7\\
\hline
\hline
\end{tabular}
\end{table}

\subsection{Additional Analysis of Group Behavior Distribution}
\label{sec:appx:group-behvaior-distrbution}
\begin{figure}[t]
    \minipage{0.48\textwidth}
    \centering
    \includegraphics[width=\linewidth]{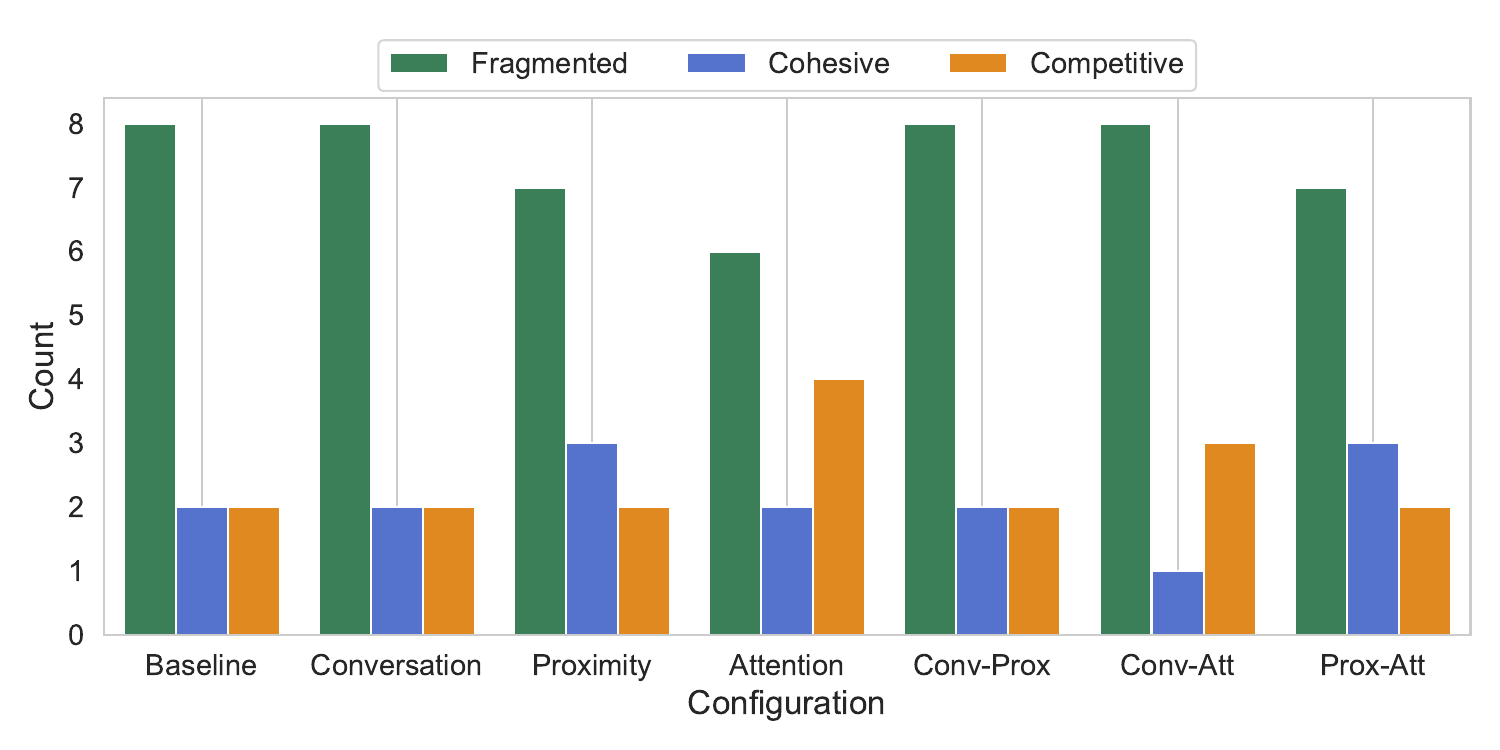}
    \vspace{-0.7cm}
    \caption{Distribution of group behavior across configurations,
    showing the overall dominance of a particular group behavior by configuration.}
    \label{fig:dist-group-level}
    \endminipage \hfill
    \minipage{0.48\textwidth}
        \centering
    \includegraphics[width=\linewidth]{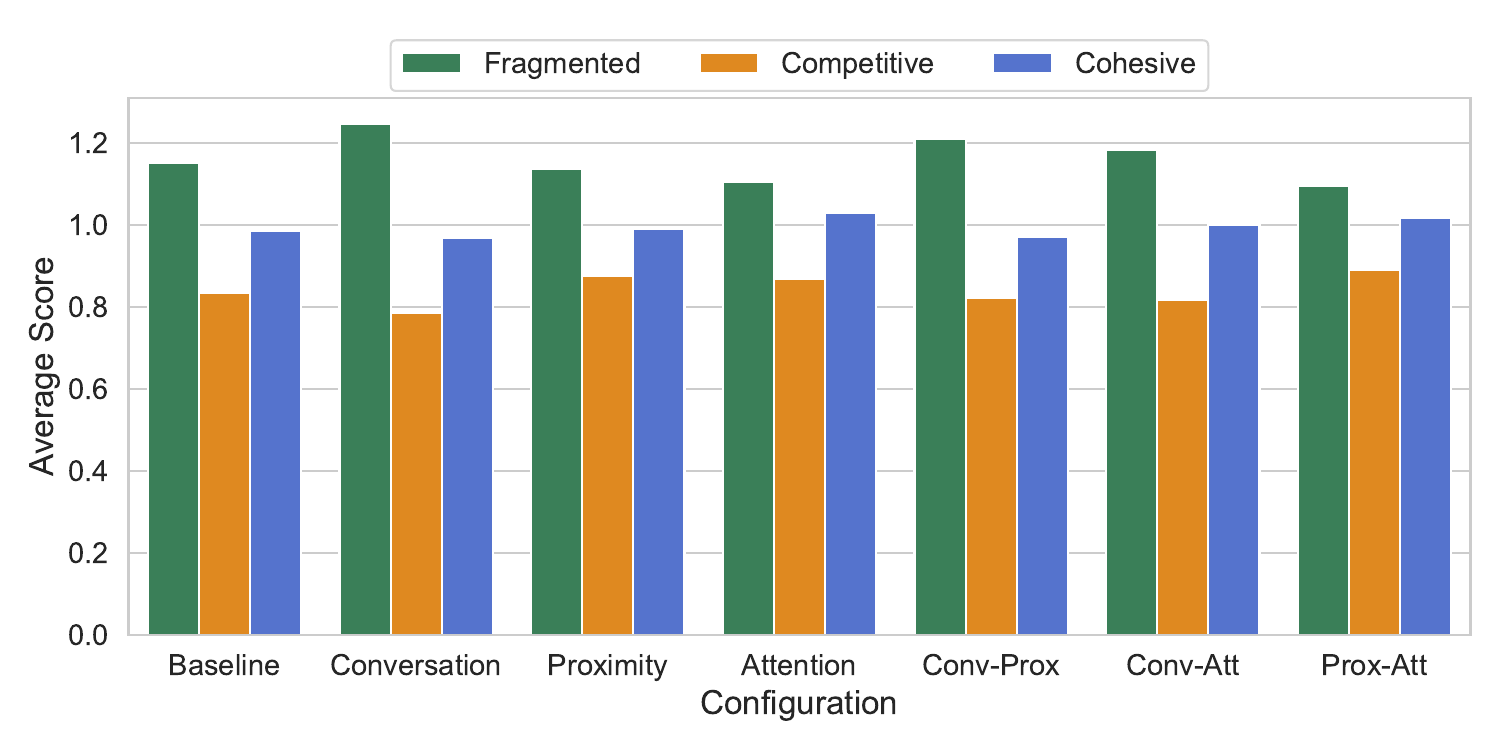}
    \vspace{-0.7cm}
    \caption{Average scores per group behavior across configurations, highlighting the mean scores for each behavior type within different configuration settings.}
    \label{fig:group-label-average-score}
    \endminipage
\end{figure}

We evaluated the impact of different sociogram weight combinations on group behavior predictions. Multiple configurations were tested, ranging from equal weighting to skewed distributions where one interaction type is emphasized. For example, the conversation-focused setup assigns 0.5 weight to conversation and 0.25 to others, while balanced setups like conversation-proximity (0.4 each, 0.2 for shared attention) explore dual-context effects. In~\autoref{fig:dist-group-level}, the label distribution shows how weighting affects dominant group behaviors. Conversation-focused setups produced more cohesive groups, while balanced configurations (e.g., Conv-Prox, Conv-Att) showed varied behaviors, supporting diverse group interactions. In~\autoref{fig:group-label-average-score}, it highlights average scores by configuration, with conversation and proximity-focused setups showing higher cohesive scores. Prox-Att configurations indicated increased competitiveness but did not dominate, suggesting that while competitiveness is measurable, it is not the primary behavior. 

\subsection{Additional \sysname Stability Analysis}\label{sec:appx:stability_analysis}
\begin{figure}[t]
  \centering
    \includegraphics[width=0.7\linewidth]{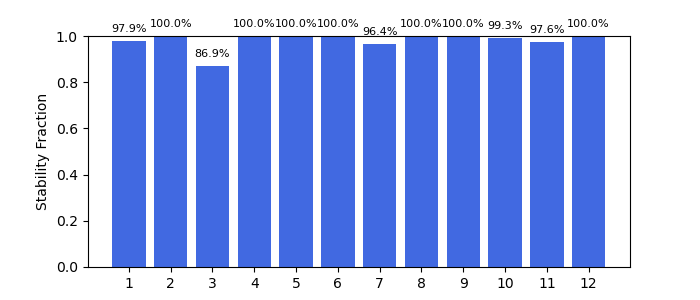}
    \vspace{-0.3cm}
   \caption{Stability Analysis Results for Group Label Assignments. The graph illustrates the robustness of the \sysname algorithm across 1000 iterations with $±5\%$ random uniform noise applied to edge weights. Bars represent the percentage of iterations where each group maintained its original label assignment. }
  \label{fig:stability}
\end{figure}

We evaluated \sysname's robustness through a stability analysis, introducing $±5\%$ random noise in sociograms via Monte Carlo simulations (1000 iterations across 12 groups). This analysis helps understand performance in real-world conditions with inherent noise and variability. 
Results in~\autoref{fig:stability} show $98.17\%$ label consistency, confirming resilience to minor perturbations, indicating \sysname algorithm's robustness to minor edge weight fluctuations, hinting reliable group behavior classification. Group 3, closest to a decision boundary, shifted to Fragmented in $13.1\%$ of cases, suggesting refinement needs for threshold tuning. This analysis validates \sysname’s reliability in classifying group behavior in MR while identifying areas for improvement in the future.

\section{Additional Limitation Details}
\label{sec:appx-limitations}
\sysname provides valuable insights into group behavior in MR collaboration, but several limitations should be addressed in future research to improve its generalizability and robustness.
\subsection{Joint Attention and Neurodivergent Collaboration}
\sysname’s reliance on shared attention as a key metric may not fully capture collaborative engagement in all populations. Individuals on the autism spectrum often exhibit different patterns of joint attention, with some struggling to maintain shared focus during social interactions~\cite{caruanaJointAttentionDifficulties2018}. While \sysname integrates multiple behavioral cues beyond joint attention, it may not accurately reflect group dynamics for neurodivergent individuals who engage in collaboration differently. Future work should explore alternative or adaptive metrics that account for diverse cognitive and behavioral differences, ensuring the framework remains inclusive and representative of broader user populations.

\subsection{Limited Sensor Modalities and Behavioral Metrics}
\sysname currently analyzes group behavior through three key sociometric features: proximity, conversation, and shared attention. While these have established links to small group collaboration, additional sensing modalities could enhance behavioral analysis. For example, facial expressions and gaze direction could offer deeper insights into emotions and engagement levels. Physiological signals such as heart rate variability and galvanic skin response could further reveal cognitive load and stress levels during collaboration. Future iterations of \sysname should incorporate multimodal sensor data to provide a more holistic view of group behavior and interaction quality.

\subsection{Group Dynamics and Social Psychology Frameworks}
This study provides a snapshot of group behavior within a controlled setting but does not capture how group dynamics evolve over time. Theories from social psychology, such as Tuckman’s stages of group development~\cite{tuckman1965developmental} and Lewin’s group dynamics~\cite{lewin1947frontiers}, suggest that collaboration patterns change as groups progress through different phases of interaction. Additionally, social phenomena such as social loafing~\cite{harkins1989social} and groupthink~\cite{turner1998twenty} can significantly impact how groups function, particularly in high-stakes or time-sensitive tasks. Future research should explore how \sysname can incorporate dynamic network analysis~\cite{carley2014ora} to track group behavior over time, capturing transitions in collaboration, leadership shifts, and social influence patterns in MR environments.

\subsection{Sample Size and Task Diversity Constraints}
The study’s findings are limited by a relatively small participant pool and a single collaborative task. While the results provide important initial insights, a larger sample size would improve statistical power and enhance the reliability of the findings. Additionally, the task type may influence observed group behaviors, as different activities require varying levels of coordination, cognitive demand, and communication styles. Future studies should include diverse tasks, such as problem-solving, brainstorming, and negotiation, to assess how \sysname adapts across different collaborative settings and challenges. This would help validate the framework’s ability to generalize beyond the specific task examined in this study.

\subsection{Qualitative Insights and Subjective Experience}
Although \sysname integrates quantitative measures of group behavior, it does not capture the nuanced subjective experiences that influence collaboration. Self-reported measures, such as participant reflections, interviews, or think-aloud protocols, could provide valuable context for interpreting observed behaviors. Understanding how individuals perceive group interactions, including their frustrations, motivations, and perceived challenges, would complement the current data-driven approach. For example, participants who scored highly on shared attention metrics might still report disengagement if their attention was forced rather than voluntary. Future work should integrate qualitative methods to refine behavioral classifications and provide richer interpretations of group behavior patterns.

By addressing these limitations, future research can improve \sysname’s ability to assess group behavior comprehensively, broadening its applicability across diverse populations, tasks, and collaborative environments.

\end{document}